\documentclass[%
 aip,
 jmp,%
 amsmath,amssymb,
 reprint,%
]{revtex4-2}

\usepackage{graphicx}
\usepackage{dcolumn}
\usepackage{bm}
\usepackage{rotating}
\usepackage{mathtools}
\usepackage{multirow}
\usepackage{hyperref}
\usepackage{tikz}
\usetikzlibrary{fit,tikzmark}

\setcitestyle{numbers,square,sort&compress}

\newcommand{\Mc}{\mathcal{M}_c}
\newcommand{\chieff}{\chi_{\rm eff}}
\newcommand{\soz}{s_{1,z}}
\newcommand{\stz}{s_{2,z}}
\newcommand{\sop}{s_{1,p}}
\newcommand{\stp}{s_{2,p}}
\newcommand{\amax}{a_{\rm max}}
\newcommand{\dilog}{\,{\rm Li}_2}

\definecolor{funcolor}{HTML}{08306B}

\begin{document}

\preprint{AIP/123-QED}

\title[]{A Thesaurus for Common Priors in Gravitational-Wave Astronomy\vspace{0.6cm}}

\author{T. Callister\,\,}
\email{tcallister@flatironinstitute.org}
\affiliation{Center for Computational Astrophysics, Flatiron Institute, New York, NY 10010, USA}

\date{\today}

\begin{abstract}
In gravitational-wave data analysis, we regularly work with a host of non-trivial prior probabilities on compact binary masses, redshifts, and spins.
We must regularly manipulate these priors, computing the implied priors on a transformed basis of parameters or reweighting posterior samples from one prior to another.
Here, I detail some common manipulations, presenting a table of Jacobians with which to transform priors between mass parametrizations, describing the conversion between source- and detector-frame priors, and deriving analytic expressions for priors on the ``effective spin'' parameters regularly invoked in gravitational-wave astronomy.
\end{abstract}

\maketitle

\vspace{0.0cm}
\section{Introduction}

\noindent
Prior probability distributions play an important role in gravitational-wave astronomy.
Non-trivial priors on compact binary masses, spins, and redshifts are introduced when measuring the properties of a given system via Bayesian parameter estimation~\cite{Christensen1998,Veitch2015,Ashton2019,RomeroShaw2020}.
And farther downstream, hierarchical analysis of the compact binary population relies crucially on being able to write down and ``undo'' parameter estimation priors to make way for a new population-informed prior that we seek to infer from the data~\cite{abbott_binary_2016,abbott_binary_2019,Mandel2019,Roulet2019,wysocki_reconstructing_2019,PhysRevD.100.043030,Vitale2020,O3a}.
\\

\noindent
We not infrequently need to manipulate these priors, determining the implicit prior on some derived quantity, or transforming from one set of priors to another that is more physically justified.
Here, I list some formulas to aid in the usage and manipulation of gravitational-wave priors.
Many of the expressions below are rather easily obtained but tiring to re-derive every time they're needed.
Others require considerable calculation and/or illuminate critical operations that are frequently mentioned in the literature but rarely presented explicitly.
\\

\noindent
The contents of this document are organized as follows:
    \begin{itemize}
    \item In Sect.~\ref{sec:mass}, I present a table of Jacobians needed to transform probability distributions between different pairs of mass parameters.
    \item In Sect.~\ref{sec:redshift}, I illustrate how priors on detector-frame masses and Euclidean distance are converted to priors on redshift and source-frame masses.
    \item Finally, in Sect.~\ref{sec:spins}, I give analytic expressions translating two common priors on compact binary spins (aligned and isotropic orientations) into their implied priors on the so-called ``effective inspiral spin'' and ``effective precessing spin'' parameters.
    \end{itemize}

\noindent
Although the results in Sect.~\ref{sec:spins} are presented without proof for brevity, I have included the sometimes-lengthy derivations of these results as separate appendices.
Also, I have provided a set of \textsc{python} functions that implement the main results of Sect.~\ref{sec:spins} at: \url{https://github.com/tcallister/effective-spin-priors}~\cite{github}.

\newpage
\section{Translating Between Mass Parameters}
\label{sec:mass}

\noindent
We need only two parameters to uniquely specify the component masses of a compact binary.
However, we regularly invoke at least six parameters: the actual component masses $m_1$ and $m_2$ (with $m_2\leq m_1$), the total mass $M=m_1+m_2$, the mass ratio $q = m_2/m_1$, the \textit{symmetric} mass ratio $\eta = m_1 m_2/(m_1+m_2)^{2} = q/(1+q)^2$, and the chirp mass $\mathcal{M}_c = \eta^{3/5} M = m_1^{3/5} m_2^{3/5} M^{-1/5}$.
We often need to transform prior densities defined on one pair of mass variables into the equivalent prior density  defined on some other pair.
We might, for example, be interested in the prior defined on the two component masses (say, in order to remove said prior during hierarchical modeling) but be given posterior samples whose prior was instead defined on chirp mass and mass ratio.
\\

{
\begin{turnpage}
\begin{table}
\caption{
\label{mass-table}
Jacobians $J = \frac{\partial(A,B)}{\partial(C,D)}$ for transforming probability densities from the mass parameter pairs labeling columns to pairs labeling rows.
Elements in the empty upper-right corner are found by inverting the inverse Jacobian from the lower-left corner. \vspace{0.5cm}}
\setlength{\tabcolsep}{5pt}
\renewcommand{\arraystretch}{2.5}
\begin{tabular}{ c  c || c c c c c c c c c c}
& & \multicolumn{10}{c}{$\left(A,B\right)$}\\
    & & $(\mathcal{M}_c,M)$ 
    & $(\mathcal{M}_c,m_1)$ 
    & $(\mathcal{M}_c,m_2)$ 
    & $(\mathcal{M}_c,q)$
    & $(M,m_1)$ 
    & $(M,m_2)$ 
    & $(M,q)$ 
    & $(m_1,m_2)$ 
    & $(m_1,q)$ 
    & $(m_2,q)$ \\
\hline
\hline
\multirow{10}{*}{$(C,D)$}
& $(\mathcal{M}_c,M)$
    & $1$
    & --
    & --
    & --
    & --
    & --
    & --
    & --
    & --
    & --
    \\
& $(\mathcal{M}_c,m_1)$
    & $\dfrac{3 m_1^2 m_2^2 (m_1-m_2)}{3m_1^3 m_2^2-\mathcal{M}_c^5}$
    & $1$
    & --
    & --
    & --
    & --
    & --
    & --
    & --
    & --
    \\
& $(\mathcal{M}_c,m_2)$ 
    & $\dfrac{3 m_1^2 m_2^2 (m_1-m_2)}{3 m_1^2 m_2^3 -\mathcal{M}_c^5}$
    & $\dfrac{\mathcal{M}_c^5 - 3 m_1^3 m_2^2}{\mathcal{M}_c^5 - 3 m_1^2 m_2^3}$
    & $1$
    & --
    & --
    & --
    & --
    & --
    & --
    & --
    \\
& $(\mathcal{M}_c,q)$
    & $\dfrac{3 M(1-q)}{5q(1+q)}$
    & $\dfrac{\mathcal{M}_c (3+2q)}{5 q^{8/5} (1+q)^{4/5}}$
    & $\dfrac{M (2+3q)}{5(1+q)^2}$
    & 1
    & --
    & --
    & --
    & --
    & --
    & --
    \\
& $(M,m_1)$
    & $\dfrac{3 \mathcal{M}_c (1-q^2)}{5Mq }$
    & $\dfrac{3+2q}{5 q^{2/5} (1+q)^{6/5}}$
    & $\dfrac{\mathcal{M}_c (2+3q)}{5 M}$
    & $\dfrac{\mathcal{M}_c}{m_1^2}$
    & $1$
    & --
    & --
    & --
    & --
    & --
    \\
& $(M,m_2)$
    & $\dfrac{3 \mathcal{M}_c (1-q^2)}{5Mq }$
    & $\dfrac{3+2q}{5 q^{2/5} (1+q)^{6/5}}$
    & $\dfrac{\mathcal{M}_c (2+3q)}{5 M}$
    & $\dfrac{\mathcal{M}_c}{m_1^2}$
    & $1$
    & $1$
    & --
    & --
    & --
    & --
    \\
& $(M,q)$  
    & $\dfrac{3 \mathcal{M}_c (1-q)}{5 q (1+q)}$
    & $\dfrac{\mathcal{M}_c (3+2q)}{5q (1+q)^2}$
    & $\dfrac{\mathcal{M}_c (2+3q)}{5 (1+q)^2}$
    & $\eta^{3/5}$
    & $\dfrac{M}{(1+q)^2}$
    & $\dfrac{M}{(1+q)^2}$
    & $1$
    & --
    & --
    & --
    \\
& $(m_1,m_2)$
    & $\dfrac{3 \mathcal{M}_c(1-q^2)}{5 M q}$
    & $\dfrac{3+2q}{5 q^{2/5} (1+q)^{6/5}}$
    & $\dfrac{\mathcal{M}_c (2+3q)}{5 M}$
    & $\dfrac{\mathcal{M}_c}{m_1^2}$
    & $1$
    & $1$
    & $\dfrac{(1+q)^2}{M}$
    & $1$
    & --
    & --
    \\
& $(m_1,q)$
    & $\dfrac{3 \mathcal{M}_c(1-q)}{5 q}$
    & $\dfrac{\mathcal{M}_c (3+2q)}{ 5q(1+q)}$
    & $\dfrac{\mathcal{M}_c (2+3q)}{5(1+q)}$
    & $\eta^{3/5}(1+q)$
    & $m_1$
    & $m_1$
    & $1+q$
    & $m_1$
    & $1$
    & --
    \\
& $(m_2, q)$
    & $\dfrac{3 \mathcal{M}_c(1-q)}{5 q^2}$
    & $\dfrac{\mathcal{M}_c (3+2q)}{5 q^2 (1+q)}$
    & $\dfrac{\mathcal{M}_c (2+3q)}{5q(1+q)}$
    & $\eta^{3/5}\frac{1+q}{q}$
    & $\dfrac{m_2}{q^2}$
    & $\dfrac{m_2}{q^2}$
    & $\dfrac{1+q}{q}$
    & $\dfrac{m_2}{q^2}$
    & $\dfrac{1}{q}$
    & $1$
    \\
\end{tabular}
\end{table}
\end{turnpage}
}

\noindent
In Table~\ref{mass-table}, I list the Jacobian factors $J = \frac{\partial(A,B)}{\partial(C,D)}$ required to transform between probability densities defined on any combination of $\{\Mc,M,m_1,m_2,q\}$, where column headings denote the pair $(A,B)$ and row headings the \textit{target} pair $(C,D)$.
Jacobians obey the convenient relation $\frac{\partial(A,B)}{\partial(C,D)} = \left(\frac{\partial(C,D)}{\partial(A,B)}\right)^{-1}$, and so any Jacobian in the blank upper-right portion of the table can be obtained by inverting the Jacobian for the inverse transformation in the lower-left portion.

\section{From the detector frame to the source frame}
\label{sec:redshift}

\noindent
Another common operation is to translate priors defined on observed \textit{detector frame} quantities into the implicit priors imposed on \textit{source frame} parameters.
In a Newtonian universe, the gravitational-wave signal received at Earth from a distant source would depend on the source's distance $D$ and component masses $m_i$.
Our universe is not Newtonian, but exhibits a non-trivial expansion history governed by general relativity.
In this context, observed gravitational-wave signals depend not on source-frame component masses $m_i$, but on the detector-frame masses (or ``redshifted masses'') $\tilde m_i = m_i(1+z)$, and similarly  on the \textit{luminosity distance}
    \begin{equation}
    D_L(z) = D_C (1+z) = (1+z) \int_0^{z'} \frac{c\,dz'}{H(z')},
    \label{eq:luminosity-d}
    \end{equation}
rather than the comoving distance $D_C$.
In Eq.~\eqref{eq:luminosity-d}, $c$ is the speed of light and $H(z)$ the Hubble parameter.
Note that we are presuming a \textit{flat} Universe; in the case of non-vanishing curvature Eq.~\eqref{eq:luminosity-d} would take a different form (see Eq.~16 of Ref.~\cite{1999astro.ph..5116H}), modifying the results below.
\\

\noindent
Parameter estimation codes like \textsc{lalinference}~\cite{Veitch2015} are typically unaware of cosmology; the component masses measured are actually the detector-frame masses (denoted $\tilde m_i$), and the source distance actually a luminosity distance.
Correspondingly, the mass priors are in fact priors $p(\tilde m_i)$ on these detector-frame quantities, which imply some non-trivial joint prior on the source-frame masses and redshift of a particular source.
Meanwhile, a seemingly innocuous prior $p(D) \propto D^2$ that is uniform in volume is in actuality uniform in ``luminosity volume'': $p(D_L) \propto D_L^2$.
\\

\noindent
In hierarchical inference of the source-frame masses and redshifts, a necessary step is the removal of this prior.
This, in turn, requires knowing the prior $p(m_1,m_2,z)$ implicitly imposed by a detector-frame prior $p(\tilde m_1,\tilde m_2,D_L)$.
Given a prior probability defined on $\{\tilde m_1,\tilde m_2,D_L\}$, the corresponding density on $\{m_1,m_2,z\}$ is
    \begin{equation}
    \begin{aligned}
    p(m_1,m_2,z)
    &= \frac{d^3 P}{dm_1 dm_2 dz} \\[3pt]
    &= \frac{d^3P}{d\tilde m_1 d\tilde m_2 dD_L}
        \left|\frac{\partial (\tilde m_1,\tilde m_2,D_L)}
        {\partial(m_1,m_2,z)}\right| \\[3pt]
    &\propto \Big( p(\tilde m_1,\tilde m_2) D_L^2\Big)
        \begin{vmatrix}
        1+z & 0 & m_1 \\
        0 &  1+z & m_2 \\
        0 & 0 & \dfrac{d D_L}{dz}
        \end{vmatrix}
        \\[3pt]
    &\propto p(\tilde m_1,\tilde m_2) \left(1+z\right)^2 D_L^2(z) \frac{d D_L}{dz}.
    \end{aligned}
    \end{equation}
Using the definitions of luminosity and comoving distances from above, we get

    \begin{equation}
    \boxed{
    p(m_1,m_2,z) \propto p(\tilde m_1, \tilde m_2) \left(1+z\right)^2 D_L^2(z)
        \left[ D_C(z) + \frac{c(1+z)}{H(z)}\right]}
    \end{equation}
    

\section{Spin magnitudes, spin components, and effective spins}
\label{sec:spins}

\noindent
Priors on the spins of compact binaries are typically written down in terms of the dimensionless spin magnitude $a \in [0,1]$ and tilt angle $t$ relative to the orbital angular momentum.
It sometimes important, though, to know the corresponding implicit prior on the actual spin components: the component $s_z = a \cos t$ parallel to the orbital angular momentum, and the component $s_p = a \sin t$ lying in the orbital plane.
We also frequently work in terms of \textit{effective spin} parameters, including the effective inspiral spin
    \begin{equation}
    \chieff = \frac{a_1 \cos t_1 + q \,a_2 \cos t_2}{1+q}
    \end{equation}
quantifying the mass-weighted average spin in the $z$-direction~\cite{Damour,Racine}, and the effective precessing spin
    \begin{equation}
    \chi_p = \max\Big[ a_1 \sin t_1, \,\left(\frac{3+4q}{4+3q}\right) q \,a_2 \sin t_2\Big].
    \end{equation}
that roughly corresponds to the degree of in-plane spins~\cite{Schmidt2015}.
\\

\noindent
In the following two subsections, I consider two common priors imposed on spin magnitudes and tilts and give the corresponding implicit priors on $s_z$, $s_p$, $\chieff$, and $\chi_p$.
The derivations of these results are at times rather involved, and so are shown separately in Appendices~\ref{sec:aligned-appendix} and \ref{sec:isotropic-appendix}.

\subsection{Uniform \& aligned component spin priors}
\label{sec:aligned-spins}

\begin{figure}
    \centering
    \includegraphics[width=\textwidth]{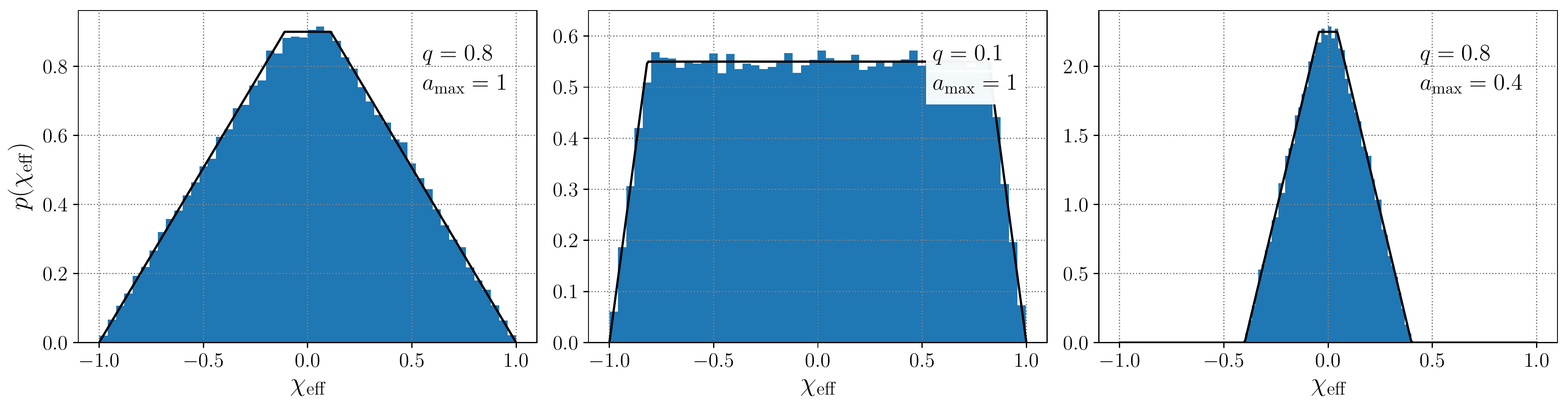}
    \caption{The $\chieff$ prior distributions constructed via random draws from aligned spin priors under several different values of $q$ and $\amax$, compared to the analytic expression in Eq.~\eqref{eq:aligned-chieff}.
    }
    \label{fig:chi_eff_aligned_demo}
\end{figure}

\noindent 
Consider a uniform distribution of component spin magnitudes, with directions assumed to be perfectly aligned with a binary's orbital angular momentum, such  that $s_p = 0$ and
    \begin{equation}
    p(\soz) = p(\stz) = \frac{1}{2\,a_{\rm max}},
    \end{equation}
defined on the interval $s_{i,z}\in [-a_{\rm max}, a_{\rm max}]$. 
Priors of this form might be used when performing parameter estimation with various families of ``aligned-spin'' waveforms, including \textsc{IMRPhenomD}~\cite{Husa2016,Khan2016} and \textsc{SEOBNRv4}~\cite{Bohe2017}.
Perhaps more importantly, aligned-spin population priors were also used in generating the injection sets~\cite{injections} employed to measure the Advanced LIGO \& Virgo selection function during the O3a observing run~\cite{O3a,gwtc2}.
\\

\noindent
Given a uniform aligned spin prior, the corresponding prior on the effective spin $\chieff$ is
    \begin{equation}
    \boxed{
    p(\chieff|q) =
        \begin{cases}
            0 &
                \Big(\chieff \leq -a_{\rm max}\,\,\mathrm{or}\,\,\chieff \geq a_{\rm max}\Big) \\[10pt]
            \dfrac{(1+q)^2 \left(a_{\rm max} - \chieff \right)}{4 \,q\, a_{\rm max}^2} & \left(\chieff > \dfrac{1-q}{1+q} \,a_{\rm max},\,\,\chieff<\amax\right) \\[10pt]
            \dfrac{(1+q)^2 \left(a_{\rm max} + \chieff \right)}{4 \,q\, a_{\rm max}^2} & \left(\chieff < -\dfrac{1-q}{1+q} \,a_{\rm max},\,\,\chieff>-\amax\right) \\[10pt]
            \dfrac{1+q}{2\,a_{\rm max}} & \big({\rm else}\big).
        \end{cases}
        }
        \label{eq:aligned-chieff}
    \end{equation}
This expression is derived in Appendix~\ref{sec:aligned-appendix}. 
Figure~\ref{fig:chi_eff_aligned_demo} compares Eq.~\eqref{eq:aligned-chieff} to $\chieff$ prior distributions constructed numerically by randomly drawing pairs of aligned spin values, subject to several different values of $q$ and $\amax$.
Note that, given its dependence on the mass ratio, Eq.~\eqref{eq:aligned-chieff} is \textit{conditional} on $q$.
If the marginal mass ratio prior $p(q)$ is known, then the \textit{joint} prior on $q$ and $\chieff$ can be expressed via the product $p(\chieff,q) = p(\chieff|q) \,p(q)$.

\subsection{Uniform \& isotropic component spin priors}
\label{sec:isotropic-spins}

\begin{figure}
    \centering
    \includegraphics[width=\textwidth]{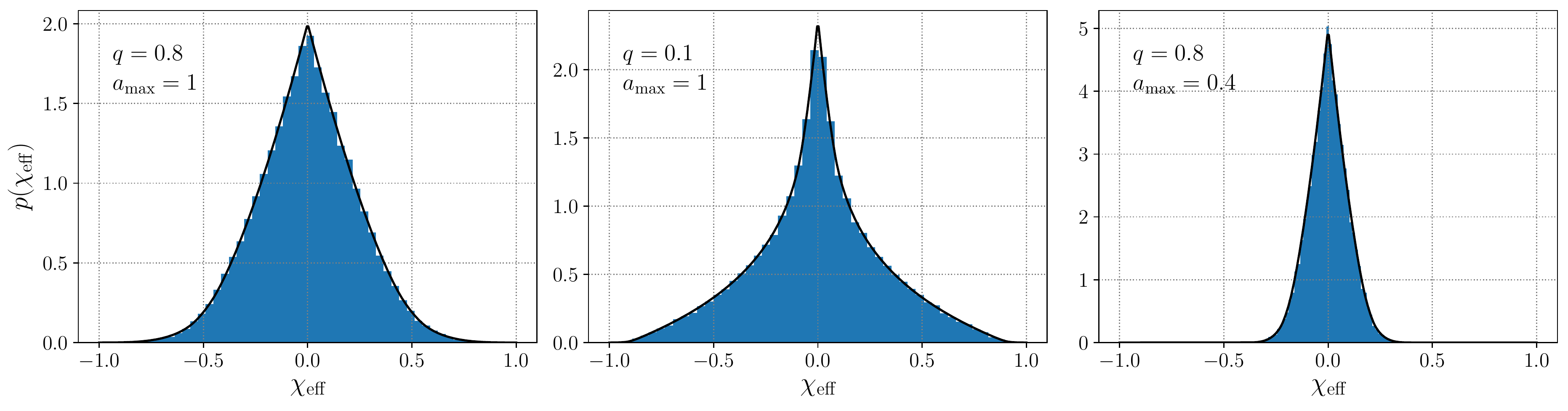}
    \caption{The $\chieff$ distributions implied by random draws from isotropic spin priors with several different values of $q$ and $\amax$, compared to the analytic expression in Eq.~\eqref{eq:final-chi-eff}.
    Together, these three examples activate every piecewise case in Eq.~\eqref{eq:final-chi-eff}.
    }
    \label{fig:chi_eff_demo}
\end{figure}

\noindent
Again consider a uniform uniform prior $p(a) = \frac{1}{a_{\rm max}}$ on component spin magnitudes on the interval $0\leq a\leq a_{\rm max}$, but now with an isotropic prior $p(\cos t) =  \frac{1}{2}$ on their direction.
The corresponding joint prior on the aligned spin component $s_z = a\cos t$ and in-plane spin $s_p = a \sin t$ is
    \begin{equation}
    \boxed{
    p(s_z,s_p) = \frac{1}{2 \amax} \frac{s_p}{s_z^2+s_p^2}.}
    \label{eq:sz-sp}
    \end{equation}
The marginal priors on $s_z$ and $s_p$ individually are
    \begin{equation}
    \boxed{
    \begin{aligned}
    p(s_z) &= \frac{1}{2 a_{\rm max}} \ln \left(\frac{a_{\rm max}}{|s_z|}\right) \\
    p(s_p) &= \frac{1}{a_{\rm max}} \cos^{-1}\left(\frac{s_p}{a_{\rm max}}\right).
    \end{aligned}
    }
    \end{equation}
\\

\noindent
The marginal prior on $\chieff$ is quite non-trivial to write down, but is most concisely expressed in the form
    \begin{equation}
    \boxed{
    p(\chieff|q) = \begin{cases}
    \mathrm{Eq.~\eqref{eq:p-xeff-0}}
        & \Big(\chieff = 0\Big)
        \\[13pt]
    \mathrm{Eq.~\eqref{eq:p-xeff-1}}
        &
        \left(
        \begin{aligned}
        |\chieff| < \amax \left(\dfrac{1-q}{1+q}\right),\,
        |\chieff| < \dfrac{q\amax}{1+q}
        \end{aligned}
        \right)
        \\[13pt]
    \mathrm{Eq.~\eqref{eq:p-xeff-2}}
            &
        \left(
        |\chieff| < \amax \left(\dfrac{1-q}{1+q}\right),\,
        |\chieff| >\dfrac{q\amax}{1+q}
        \right)
        \\[13pt]
    \mathrm{Eq.~\eqref{eq:p-xeff-3}}
        &
        \left(
        |\chieff| > \amax \left(\dfrac{1-q}{1+q}\right),\,|\chieff| < \dfrac{q\amax }{1+q}
        \right)
        \\[13pt]
    \mathrm{Eq.~\eqref{eq:p-xeff-4}}
        &
        \left(
        |\chieff| > \amax \left(\dfrac{1-q}{1+q}\right),\,|\chieff| < \dfrac{\amax}{1+q},\,|\chieff| > \dfrac{q\amax}{1+q}
        \right)
        \\[13pt]
    \mathrm{Eq.~\eqref{eq:p-xeff-5}}
        &
        \left(
        |\chieff| > \dfrac{\amax}{1+q},\, |\chieff| < \amax
        \right)
        \\[15pt]
    0 & \Big(|\chieff|\geq \amax\Big)
    \end{cases}
    }
    \label{eq:final-chi-eff}
    \end{equation}
\\

\begin{figure}
    \centering
    \includegraphics[width=\textwidth]{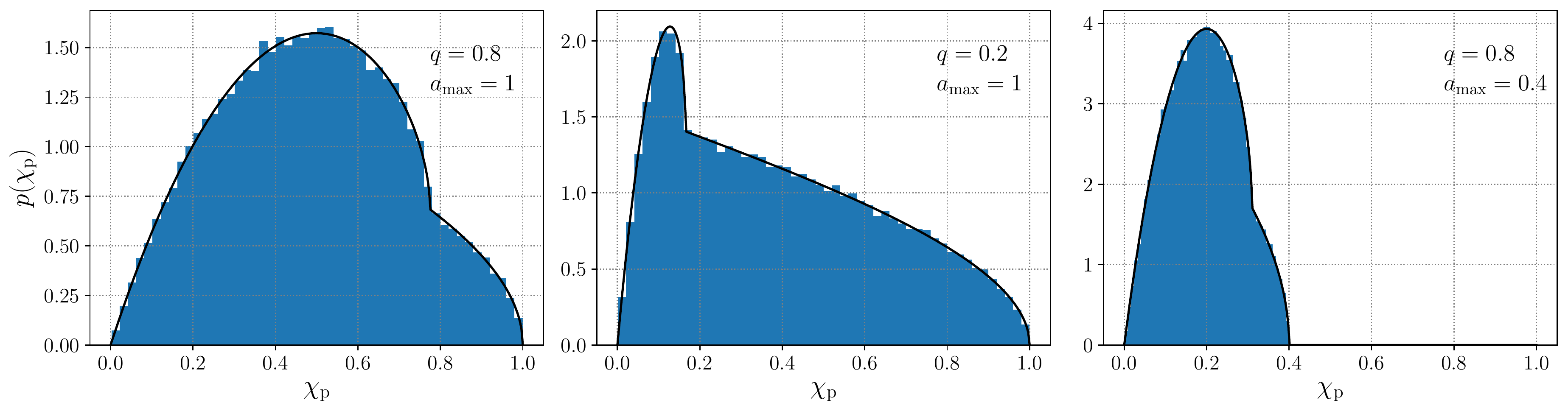}
    \caption{The $\chi_p$ distributions implied by random draws from isotropic spin priors with several different values of $q$ and $\amax$, compared to the analytic expression in Eq.~\eqref{eq:chip}.
    }
    \label{fig:chi_p_demo}
\end{figure}

\noindent
The marginal prior on $\chi_p$, meanwhile, is given by
    \begin{equation}
    \boxed{
    p(\chi_p|q) = \mathrm{Eq.~\eqref{eq:chip-1}} \, + \,\mathrm{Eq.~\eqref{eq:chip-2}}.}
    \label{eq:chip}
    \end{equation}
The derivation of Eqs.~\eqref{eq:sz-sp}-\eqref{eq:chip} is given in Appendix~\ref{sec:isotropic-appendix}, which  also contains the full expressions referenced here in Eq.~\eqref{eq:final-chi-eff} and Eq.~\eqref{eq:chip}
\\

\noindent
Figures \ref{fig:chi_eff_demo} and \ref{fig:chi_p_demo} compare these analytic (if daunting) expressions to the $\chieff$ and $\chi_p$ distributions constructed numerically by drawing random spin magnitudes and tilts from a uniform and isotropic prior.
The various combinations of $q$ and $\amax$ are chosen such that, across the three examples in Fig.~\ref{fig:chi_eff_demo}, we encounter each piecewise case in Eq.~\eqref{eq:final-chi-eff}.
\\

\noindent
Code that implements Eqs.~\eqref{eq:final-chi-eff} and \eqref{eq:chip} is available at the \textsc{github} link in Ref.~\cite{github}.
Two additional notes concerning the implementation of Eq.~\eqref{eq:final-chi-eff}:
First, as mentioned below in Appendix~\ref{sec:isotropic-appendix}, Eq.~\eqref{eq:final-chi-eff} depends on a special function called the ``dilogarithm'' or ``Spence's function''~\cite{spence}.
\textsc{Mathematica}~\cite{Mathematica} and \textsc{scipy}~\cite{scipy} adopt different conventions in their implementation of this function.
In Appendix~\ref{sec:isotropic-appendix}, I follow \textsc{Mathematica}'s convention, such that the quantity I denote $\dilog(z)$ is equivalent to \textsc{Mathematica}'s {\texttt{PolyLog[2,z]}}.
When translated to \textsc{python}, the corresponding quantity is obtained by calling \texttt{scipy.special.spence(1-z)}.
Second, a very careful reader or user will notice that Eq.~\eqref{eq:final-chi-eff} is undefined on the \textit{boundaries} between cases.
Rather than consider every possible boundary (of which there are many) as a separate edge case, I instead deal with boundary cases simply by averaging the nearby values $p(\chieff+\epsilon|q)$ and $p(\chieff-\epsilon|q)$, for a small offset $\epsilon$.
Since $p(\chieff|q)$ is quite smooth everywhere except the origin, this gives an accurate estimate of the prior at any $\chieff$ sitting on a boundary between two cases.
\\

\noindent
\section*{Acknowledgements}
\noindent Deep thanks to Will Farr, Colm Talbot, and Daniel Wysocki for their valuable thoughts and feedback on these notes.

\bibliography{priors.bib}

\providecommand{\noopsort}[1]{}\providecommand{\singleletter}[1]{#1}%
\begin{thebibliography}{25}%
\makeatletter
\providecommand \@ifxundefined [1]{%
 \@ifx{#1\undefined}
}%
\providecommand \@ifnum [1]{%
 \ifnum #1\expandafter \@firstoftwo
 \else \expandafter \@secondoftwo
 \fi
}%
\providecommand \@ifx [1]{%
 \ifx #1\expandafter \@firstoftwo
 \else \expandafter \@secondoftwo
 \fi
}%
\providecommand \natexlab [1]{#1}%
\providecommand \enquote  [1]{``#1''}%
\providecommand \bibnamefont  [1]{#1}%
\providecommand \bibfnamefont [1]{#1}%
\providecommand \citenamefont [1]{#1}%
\providecommand \href@noop [0]{\@secondoftwo}%
\providecommand \href [0]{\begingroup \@sanitize@url \@href}%
\providecommand \@href[1]{\@@startlink{#1}\@@href}%
\providecommand \@@href[1]{\endgroup#1\@@endlink}%
\providecommand \@sanitize@url [0]{\catcode `\\12\catcode `\$12\catcode
  `\&12\catcode `\#12\catcode `\^12\catcode `\_12\catcode `\%12\relax}%
\providecommand \@@startlink[1]{}%
\providecommand \@@endlink[0]{}%
\providecommand \url  [0]{\begingroup\@sanitize@url \@url }%
\providecommand \@url [1]{\endgroup\@href {#1}{\urlprefix }}%
\providecommand \urlprefix  [0]{URL }%
\providecommand \Eprint [0]{\href }%
\providecommand \doibase [0]{https://doi.org/}%
\providecommand \selectlanguage [0]{\@gobble}%
\providecommand \bibinfo  [0]{\@secondoftwo}%
\providecommand \bibfield  [0]{\@secondoftwo}%
\providecommand \translation [1]{[#1]}%
\providecommand \BibitemOpen [0]{}%
\providecommand \bibitemStop [0]{}%
\providecommand \bibitemNoStop [0]{.\EOS\space}%
\providecommand \EOS [0]{\spacefactor3000\relax}%
\providecommand \BibitemShut  [1]{\csname bibitem#1\endcsname}%
\let\auto@bib@innerbib\@empty
\bibitem [{\citenamefont {{Christensen}}\ and\ \citenamefont
  {{Meyer}}(1998)}]{Christensen1998}%
  \BibitemOpen
  \bibfield  {author} {\bibinfo {author} {\bibfnamefont {N.}~\bibnamefont
  {{Christensen}}}\ and\ \bibinfo {author} {\bibfnamefont {R.}~\bibnamefont
  {{Meyer}}},\ }\bibfield  {title} {\enquote {\bibinfo {title} {{Markov chain
  Monte Carlo methods for Bayesian gravitational radiation data analysis}},}\
  }\href {https://doi.org/10.1103/PhysRevD.58.082001} {\bibfield  {journal}
  {\bibinfo  {journal} {\prd}\ }\textbf {\bibinfo {volume} {58}},\ \bibinfo
  {eid} {082001} (\bibinfo {year} {1998})}\BibitemShut {NoStop}%
\bibitem [{\citenamefont {{Veitch}}\ \emph {et~al.}(2015)\citenamefont
  {{Veitch}}, \citenamefont {{Raymond}}, \citenamefont {{Farr}}, \citenamefont
  {{Farr}}, \citenamefont {{Graff}} \emph {et~al.}}]{Veitch2015}%
  \BibitemOpen
  \bibfield  {author} {\bibinfo {author} {\bibfnamefont {J.}~\bibnamefont
  {{Veitch}}}, \bibinfo {author} {\bibfnamefont {V.}~\bibnamefont {{Raymond}}},
  \bibinfo {author} {\bibfnamefont {B.}~\bibnamefont {{Farr}}}, \bibinfo
  {author} {\bibfnamefont {W.}~\bibnamefont {{Farr}}}, \bibinfo {author}
  {\bibfnamefont {P.}~\bibnamefont {{Graff}}}, \emph {et~al.},\ }\bibfield
  {title} {\enquote {\bibinfo {title} {{Parameter estimation for compact
  binaries with ground-based gravitational-wave observations using the
  LALInference software library}},}\ }\href
  {https://doi.org/10.1103/PhysRevD.91.042003} {\bibfield  {journal} {\bibinfo
  {journal} {\prd}\ }\textbf {\bibinfo {volume} {91}},\ \bibinfo {eid} {042003}
  (\bibinfo {year} {2015})},\ \Eprint {https://arxiv.org/abs/1409.7215}
  {arXiv:1409.7215} \BibitemShut {NoStop}%
\bibitem [{\citenamefont {{Ashton}}\ \emph {et~al.}(2019)\citenamefont
  {{Ashton}}, \citenamefont {{H{\"u}bner}}, \citenamefont {{Lasky}},
  \citenamefont {{Talbot}}, \citenamefont {{Ackley}} \emph
  {et~al.}}]{Ashton2019}%
  \BibitemOpen
  \bibfield  {author} {\bibinfo {author} {\bibfnamefont {G.}~\bibnamefont
  {{Ashton}}}, \bibinfo {author} {\bibfnamefont {M.}~\bibnamefont
  {{H{\"u}bner}}}, \bibinfo {author} {\bibfnamefont {P.~D.}\ \bibnamefont
  {{Lasky}}}, \bibinfo {author} {\bibfnamefont {C.}~\bibnamefont {{Talbot}}},
  \bibinfo {author} {\bibfnamefont {K.}~\bibnamefont {{Ackley}}}, \emph
  {et~al.},\ }\bibfield  {title} {\enquote {\bibinfo {title} {{BILBY: A
  User-friendly Bayesian Inference Library for Gravitational-wave
  Astronomy}},}\ }\href {https://doi.org/10.3847/1538-4365/ab06fc} {\bibfield
  {journal} {\bibinfo  {journal} {Astrophys. J. Suppl. Ser.}\ }\textbf
  {\bibinfo {volume} {241}},\ \bibinfo {eid} {27} (\bibinfo {year} {2019})},\
  \Eprint {https://arxiv.org/abs/1811.02042} {arXiv:1811.02042} \BibitemShut
  {NoStop}%
\bibitem [{\citenamefont {{Romero-Shaw}}\ \emph {et~al.}(2020)\citenamefont
  {{Romero-Shaw}}, \citenamefont {{Talbot}}, \citenamefont {{Biscoveanu}},
  \citenamefont {{D'Emilio}}, \citenamefont {{Ashton}} \emph
  {et~al.}}]{RomeroShaw2020}%
  \BibitemOpen
  \bibfield  {author} {\bibinfo {author} {\bibfnamefont {I.~M.}\ \bibnamefont
  {{Romero-Shaw}}}, \bibinfo {author} {\bibfnamefont {C.}~\bibnamefont
  {{Talbot}}}, \bibinfo {author} {\bibfnamefont {S.}~\bibnamefont
  {{Biscoveanu}}}, \bibinfo {author} {\bibfnamefont {V.}~\bibnamefont
  {{D'Emilio}}}, \bibinfo {author} {\bibfnamefont {G.}~\bibnamefont
  {{Ashton}}}, \emph {et~al.},\ }\bibfield  {title} {\enquote {\bibinfo {title}
  {{Bayesian inference for compact binary coalescences with BILBY: validation
  and application to the first LIGO-Virgo gravitational-wave transient
  catalogue}},}\ }\href {https://doi.org/10.1093/mnras/staa2850} {\bibfield
  {journal} {\bibinfo  {journal} {Mon. Not. R. Astron. Soc.}\ }\textbf
  {\bibinfo {volume} {499}},\ \bibinfo {pages} {3295--3319} (\bibinfo {year}
  {2020})},\ \Eprint {https://arxiv.org/abs/2006.00714} {arXiv:2006.00714}
  \BibitemShut {NoStop}%
\bibitem [{\citenamefont {{The LIGO Scientific Collaboration and the Virgo
  Collaboration}}(2016)}]{abbott_binary_2016}%
  \BibitemOpen
  \bibfield  {author} {\bibinfo {author} {\bibnamefont {{The LIGO Scientific
  Collaboration and the Virgo Collaboration}}},\ }\bibfield  {title} {\enquote
  {\bibinfo {title} {Binary {Black} {Hole} {Mergers} in the {First} {Advanced}
  {LIGO} {Observing} {Run}},}\ }\href
  {https://doi.org/10.1103/PhysRevX.6.041015} {\bibfield  {journal} {\bibinfo
  {journal} {Phys. Rev. X}\ }\textbf {\bibinfo {volume} {6}},\ \bibinfo {pages}
  {041015--041015} (\bibinfo {year} {2016})},\ \Eprint
  {https://arxiv.org/abs/1606.04856} {arXiv:1606.04856} \BibitemShut {NoStop}%
\bibitem [{\citenamefont {{The LIGO Scientific Collaboration and the Virgo
  Collaboration}}(2019)}]{abbott_binary_2019}%
  \BibitemOpen
  \bibfield  {author} {\bibinfo {author} {\bibnamefont {{The LIGO Scientific
  Collaboration and the Virgo Collaboration}}},\ }\bibfield  {title} {\enquote
  {\bibinfo {title} {Binary {Black} {Hole} {Population} {Properties} {Inferred}
  from the {First} and {Second} {Observing} {Runs} of {Advanced} {LIGO} and
  {Advanced} {Virgo}},}\ }\href {https://doi.org/10.3847/2041-8213/ab3800}
  {\bibfield  {journal} {\bibinfo  {journal} {Astrophys. J.}\ }\textbf
  {\bibinfo {volume} {882}},\ \bibinfo {pages} {L24} (\bibinfo {year}
  {2019})},\ \Eprint {https://arxiv.org/abs/1811.12940} {arXiv:1811.12940}
  \BibitemShut {NoStop}%
\bibitem [{\citenamefont {{Mandel}}, \citenamefont {{Farr}},\ and\
  \citenamefont {{Gair}}(2019)}]{Mandel2019}%
  \BibitemOpen
  \bibfield  {author} {\bibinfo {author} {\bibfnamefont {I.}~\bibnamefont
  {{Mandel}}}, \bibinfo {author} {\bibfnamefont {W.~M.}\ \bibnamefont
  {{Farr}}},\ and\ \bibinfo {author} {\bibfnamefont {J.~R.}\ \bibnamefont
  {{Gair}}},\ }\bibfield  {title} {\enquote {\bibinfo {title} {{Extracting
  distribution parameters from multiple uncertain observations with selection
  biases}},}\ }\href {https://doi.org/10.1093/mnras/stz896} {\bibfield
  {journal} {\bibinfo  {journal} {Mon. Not. R. Astron. Soc.}\ }\textbf
  {\bibinfo {volume} {486}},\ \bibinfo {pages} {1086--1093} (\bibinfo {year}
  {2019})},\ \Eprint {https://arxiv.org/abs/1809.02063} {arXiv:1809.02063}
  \BibitemShut {NoStop}%
\bibitem [{\citenamefont {{Roulet}}\ and\ \citenamefont
  {{Zaldarriaga}}(2019)}]{Roulet2019}%
  \BibitemOpen
  \bibfield  {author} {\bibinfo {author} {\bibfnamefont {J.}~\bibnamefont
  {{Roulet}}}\ and\ \bibinfo {author} {\bibfnamefont {M.}~\bibnamefont
  {{Zaldarriaga}}},\ }\bibfield  {title} {\enquote {\bibinfo {title}
  {{Constraints on binary black hole populations from LIGO-Virgo
  detections}},}\ }\href {https://doi.org/10.1093/mnras/stz226} {\bibfield
  {journal} {\bibinfo  {journal} {Mon. Not. R. Astron. Soc.}\ }\textbf
  {\bibinfo {volume} {484}},\ \bibinfo {pages} {4216--4229} (\bibinfo {year}
  {2019})},\ \Eprint {https://arxiv.org/abs/1806.10610} {arXiv:1806.10610}
  \BibitemShut {NoStop}%
\bibitem [{\citenamefont {Wysocki}, \citenamefont {Lange},\ and\ \citenamefont
  {O’Shaughnessy}(2019)}]{wysocki_reconstructing_2019}%
  \BibitemOpen
  \bibfield  {author} {\bibinfo {author} {\bibfnamefont {D.}~\bibnamefont
  {Wysocki}}, \bibinfo {author} {\bibfnamefont {J.}~\bibnamefont {Lange}},\
  and\ \bibinfo {author} {\bibfnamefont {R.}~\bibnamefont {O’Shaughnessy}},\
  }\bibfield  {title} {\enquote {\bibinfo {title} {Reconstructing
  phenomenological distributions of compact binaries via gravitational wave
  observations},}\ }\href {https://doi.org/10.1103/PhysRevD.100.043012}
  {\bibfield  {journal} {\bibinfo  {journal} {Phys. Rev. D}\ }\textbf {\bibinfo
  {volume} {100}},\ \bibinfo {pages} {043012} (\bibinfo {year} {2019})},\
  \Eprint {https://arxiv.org/abs/1805.06442} {arXiv:1805.06442} \BibitemShut
  {NoStop}%
\bibitem [{\citenamefont {Talbot}\ \emph {et~al.}(2019)\citenamefont {Talbot},
  \citenamefont {Smith}, \citenamefont {Thrane},\ and\ \citenamefont
  {Poole}}]{PhysRevD.100.043030}%
  \BibitemOpen
  \bibfield  {author} {\bibinfo {author} {\bibfnamefont {C.}~\bibnamefont
  {Talbot}}, \bibinfo {author} {\bibfnamefont {R.}~\bibnamefont {Smith}},
  \bibinfo {author} {\bibfnamefont {E.}~\bibnamefont {Thrane}},\ and\ \bibinfo
  {author} {\bibfnamefont {G.~B.}\ \bibnamefont {Poole}},\ }\bibfield  {title}
  {\enquote {\bibinfo {title} {Parallelized inference for gravitational-wave
  astronomy},}\ }\href {https://doi.org/10.1103/PhysRevD.100.043030} {\bibfield
   {journal} {\bibinfo  {journal} {Phys. Rev. D}\ }\textbf {\bibinfo {volume}
  {100}},\ \bibinfo {pages} {043030} (\bibinfo {year} {2019})},\ \Eprint
  {https://arxiv.org/abs/1904.02863} {arXiv:1904.02863} \BibitemShut {NoStop}%
\bibitem [{\citenamefont {{Vitale}}\ \emph {et~al.}(2020)\citenamefont
  {{Vitale}}, \citenamefont {{Gerosa}}, \citenamefont {{Farr}},\ and\
  \citenamefont {{Taylor}}}]{Vitale2020}%
  \BibitemOpen
  \bibfield  {author} {\bibinfo {author} {\bibfnamefont {S.}~\bibnamefont
  {{Vitale}}}, \bibinfo {author} {\bibfnamefont {D.}~\bibnamefont {{Gerosa}}},
  \bibinfo {author} {\bibfnamefont {W.~M.}\ \bibnamefont {{Farr}}},\ and\
  \bibinfo {author} {\bibfnamefont {S.~R.}\ \bibnamefont {{Taylor}}},\
  }\bibfield  {title} {\enquote {\bibinfo {title} {{Inferring the properties of
  a population of compact binaries in presence of selection effects}},}\
  }\href@noop {} {\  (\bibinfo {year} {2020})},\ \Eprint
  {https://arxiv.org/abs/2007.05579} {arXiv:2007.05579} \BibitemShut {NoStop}%
\bibitem [{\citenamefont {{The LIGO Scientific Collaboration and the Virgo
  Collaboration}}(2020{\natexlab{a}})}]{O3a}%
  \BibitemOpen
  \bibfield  {author} {\bibinfo {author} {\bibnamefont {{The LIGO Scientific
  Collaboration and the Virgo Collaboration}}},\ }\bibfield  {title} {\enquote
  {\bibinfo {title} {Population {Properties} of {Compact} {Objects} from the
  {Second} {LIGO}-{Virgo} {Gravitational}-{Wave} {Transient} {Catalog}},}\
  }\href {http://arxiv.org/abs/2010.14533} {\  (\bibinfo {year}
  {2020}{\natexlab{a}})},\ \bibinfo {note} {arXiv:2010.14533}\BibitemShut
  {NoStop}%
\bibitem [{\citenamefont {Callister}()}]{github}%
  \BibitemOpen
  \bibfield  {author} {\bibinfo {author} {\bibfnamefont {T.}~\bibnamefont
  {Callister}},\ }\href@noop {} {\enquote {\bibinfo {title} {Effective spin
  priors},}\ }\bibinfo {howpublished}
  {\url{https://github.com/tcallister/effective-spin-priors}},\ \bibinfo {note}
  {accessed: 2021-03-25}\BibitemShut {NoStop}%
\bibitem [{\citenamefont {{Hogg}}(1999)}]{1999astro.ph..5116H}%
  \BibitemOpen
  \bibfield  {author} {\bibinfo {author} {\bibfnamefont {D.~W.}\ \bibnamefont
  {{Hogg}}},\ }\bibfield  {title} {\enquote {\bibinfo {title} {{Distance
  measures in cosmology}},}\ }\href@noop {} {\  (\bibinfo {year} {1999})},\
  \Eprint {https://arxiv.org/abs/astro-ph/9905116} {arXiv:astro-ph/9905116}
  \BibitemShut {NoStop}%
\bibitem [{\citenamefont {Damour}(2001)}]{Damour}%
  \BibitemOpen
  \bibfield  {author} {\bibinfo {author} {\bibfnamefont {T.}~\bibnamefont
  {Damour}},\ }\bibfield  {title} {\enquote {\bibinfo {title} {Coalescence of
  two spinning black holes: An effective one-body approach},}\ }\href
  {https://doi.org/10.1103/PhysRevD.64.124013} {\bibfield  {journal} {\bibinfo
  {journal} {Phys. Rev. D}\ }\textbf {\bibinfo {volume} {64}},\ \bibinfo
  {pages} {124013} (\bibinfo {year} {2001})},\ \Eprint
  {https://arxiv.org/abs/gr-qc/0103018} {arXiv:gr-qc/0103018} \BibitemShut
  {NoStop}%
\bibitem [{\citenamefont {{Racine}}(2008)}]{Racine}%
  \BibitemOpen
  \bibfield  {author} {\bibinfo {author} {\bibfnamefont {{\'E}.}~\bibnamefont
  {{Racine}}},\ }\bibfield  {title} {\enquote {\bibinfo {title} {{Analysis of
  spin precession in binary black hole systems including quadrupole-monopole
  interaction}},}\ }\href {https://doi.org/10.1103/PhysRevD.78.044021}
  {\bibfield  {journal} {\bibinfo  {journal} {\prd}\ }\textbf {\bibinfo
  {volume} {78}},\ \bibinfo {eid} {044021} (\bibinfo {year} {2008})},\ \Eprint
  {https://arxiv.org/abs/0803.1820} {arXiv:0803.1820} \BibitemShut {NoStop}%
\bibitem [{\citenamefont {Schmidt}, \citenamefont {Ohme},\ and\ \citenamefont
  {Hannam}(2015)}]{Schmidt2015}%
  \BibitemOpen
  \bibfield  {author} {\bibinfo {author} {\bibfnamefont {P.}~\bibnamefont
  {Schmidt}}, \bibinfo {author} {\bibfnamefont {F.}~\bibnamefont {Ohme}},\ and\
  \bibinfo {author} {\bibfnamefont {M.}~\bibnamefont {Hannam}},\ }\bibfield
  {title} {\enquote {\bibinfo {title} {Towards models of gravitational
  waveforms from generic binaries: Ii. modelling precession effects with a
  single effective precession parameter},}\ }\href
  {https://doi.org/10.1103/PhysRevD.91.024043} {\bibfield  {journal} {\bibinfo
  {journal} {Phys. Rev. D}\ }\textbf {\bibinfo {volume} {91}},\ \bibinfo
  {pages} {024043} (\bibinfo {year} {2015})},\ \Eprint
  {https://arxiv.org/abs/1408.1810} {arXiv:1408.1810} \BibitemShut {NoStop}%
\bibitem [{\citenamefont {{Husa}}\ \emph {et~al.}(2016)\citenamefont {{Husa}},
  \citenamefont {{Khan}}, \citenamefont {{Hannam}}, \citenamefont
  {{P{\"u}rrer}}, \citenamefont {{Ohme}} \emph {et~al.}}]{Husa2016}%
  \BibitemOpen
  \bibfield  {author} {\bibinfo {author} {\bibfnamefont {S.}~\bibnamefont
  {{Husa}}}, \bibinfo {author} {\bibfnamefont {S.}~\bibnamefont {{Khan}}},
  \bibinfo {author} {\bibfnamefont {M.}~\bibnamefont {{Hannam}}}, \bibinfo
  {author} {\bibfnamefont {M.}~\bibnamefont {{P{\"u}rrer}}}, \bibinfo {author}
  {\bibfnamefont {F.}~\bibnamefont {{Ohme}}}, \emph {et~al.},\ }\bibfield
  {title} {\enquote {\bibinfo {title} {{Frequency-domain gravitational waves
  from nonprecessing black-hole binaries. I. New numerical waveforms and
  anatomy of the signal}},}\ }\href
  {https://doi.org/10.1103/PhysRevD.93.044006} {\bibfield  {journal} {\bibinfo
  {journal} {\prd}\ }\textbf {\bibinfo {volume} {93}},\ \bibinfo {eid} {044006}
  (\bibinfo {year} {2016})},\ \Eprint {https://arxiv.org/abs/1508.07250}
  {arXiv:1508.07250} \BibitemShut {NoStop}%
\bibitem [{\citenamefont {{Khan}}\ \emph {et~al.}(2016)\citenamefont {{Khan}},
  \citenamefont {{Husa}}, \citenamefont {{Hannam}}, \citenamefont {{Ohme}},
  \citenamefont {{P{\"u}rrer}} \emph {et~al.}}]{Khan2016}%
  \BibitemOpen
  \bibfield  {author} {\bibinfo {author} {\bibfnamefont {S.}~\bibnamefont
  {{Khan}}}, \bibinfo {author} {\bibfnamefont {S.}~\bibnamefont {{Husa}}},
  \bibinfo {author} {\bibfnamefont {M.}~\bibnamefont {{Hannam}}}, \bibinfo
  {author} {\bibfnamefont {F.}~\bibnamefont {{Ohme}}}, \bibinfo {author}
  {\bibfnamefont {M.}~\bibnamefont {{P{\"u}rrer}}}, \emph {et~al.},\ }\bibfield
   {title} {\enquote {\bibinfo {title} {{Frequency-domain gravitational waves
  from nonprecessing black-hole binaries. II. A phenomenological model for the
  advanced detector era}},}\ }\href
  {https://doi.org/10.1103/PhysRevD.93.044007} {\bibfield  {journal} {\bibinfo
  {journal} {\prd}\ }\textbf {\bibinfo {volume} {93}},\ \bibinfo {eid} {044007}
  (\bibinfo {year} {2016})},\ \Eprint {https://arxiv.org/abs/1508.07253}
  {arXiv:1508.07253} \BibitemShut {NoStop}%
\bibitem [{\citenamefont {{Boh{\'e}}}\ \emph {et~al.}(2017)\citenamefont
  {{Boh{\'e}}}, \citenamefont {{Shao}}, \citenamefont {{Taracchini}},
  \citenamefont {{Buonanno}}, \citenamefont {{Babak}} \emph
  {et~al.}}]{Bohe2017}%
  \BibitemOpen
  \bibfield  {author} {\bibinfo {author} {\bibfnamefont {A.}~\bibnamefont
  {{Boh{\'e}}}}, \bibinfo {author} {\bibfnamefont {L.}~\bibnamefont {{Shao}}},
  \bibinfo {author} {\bibfnamefont {A.}~\bibnamefont {{Taracchini}}}, \bibinfo
  {author} {\bibfnamefont {A.}~\bibnamefont {{Buonanno}}}, \bibinfo {author}
  {\bibfnamefont {S.}~\bibnamefont {{Babak}}}, \emph {et~al.},\ }\bibfield
  {title} {\enquote {\bibinfo {title} {{Improved effective-one-body model of
  spinning, nonprecessing binary black holes for the era of gravitational-wave
  astrophysics with advanced detectors}},}\ }\href
  {https://doi.org/10.1103/PhysRevD.95.044028} {\bibfield  {journal} {\bibinfo
  {journal} {\prd}\ }\textbf {\bibinfo {volume} {95}},\ \bibinfo {eid} {044028}
  (\bibinfo {year} {2017})},\ \Eprint {https://arxiv.org/abs/1611.03703}
  {arXiv:1611.03703} \BibitemShut {NoStop}%
\bibitem [{\citenamefont {{The LIGO Scientific Collaboration and the Virgo
  Collaboration}}()}]{injections}%
  \BibitemOpen
  \bibfield  {author} {\bibinfo {author} {\bibnamefont {{The LIGO Scientific
  Collaboration and the Virgo Collaboration}}},\ }\href@noop {} {\enquote
  {\bibinfo {title} {{GWTC-2} data release: {Sensitivity} of matched filter
  searches to binary black hole merger populations},}\ }\bibinfo {howpublished}
  {\url{https://dcc.ligo.org/LIGO-P2000217/public}},\ \bibinfo {note}
  {accessed: 2021-04-01}\BibitemShut {NoStop}%
\bibitem [{\citenamefont {{The LIGO Scientific Collaboration and the Virgo
  Collaboration}}(2020{\natexlab{b}})}]{gwtc2}%
  \BibitemOpen
  \bibfield  {author} {\bibinfo {author} {\bibnamefont {{The LIGO Scientific
  Collaboration and the Virgo Collaboration}}},\ }\bibfield  {title} {\enquote
  {\bibinfo {title} {{GWTC}-2: {Compact} {Binary} {Coalescences} {Observed} by
  {LIGO} and {Virgo} {During} the {First} {Half} of the {Third} {Observing}
  {Run}},}\ }\href {http://arxiv.org/abs/2010.14527} {\  (\bibinfo {year}
  {2020}{\natexlab{b}})},\ \bibinfo {note} {arXiv: 2010.14527}\BibitemShut
  {NoStop}%
\bibitem [{spe()}]{spence}%
  \BibitemOpen
  \href@noop {} {\enquote {\bibinfo {title} {Spence's function},}\ }\bibinfo
  {howpublished} {\url{https://en.wikipedia.org/wiki/Spence\%27s_function}},\
  \bibinfo {note} {accessed: 2021-04-01}\BibitemShut {NoStop}%
\bibitem [{\citenamefont {{Wolfram Research{,} Inc.}}()}]{Mathematica}%
  \BibitemOpen
  \bibfield  {author} {\bibinfo {author} {\bibnamefont {{Wolfram Research{,}
  Inc.}}},\ }\href {https://www.wolfram.com/mathematica} {\enquote {\bibinfo
  {title} {Mathematica, {V}ersion 12.2},}\ }\bibinfo {note} {Champaign, IL,
  2020}\BibitemShut {NoStop}%
\bibitem [{\citenamefont {Virtanen}\ \emph {et~al.}(2020)\citenamefont
  {Virtanen}, \citenamefont {Gommers}, \citenamefont {Oliphant}, \citenamefont
  {Haberland}, \citenamefont {Reddy} \emph {et~al.}}]{scipy}%
  \BibitemOpen
  \bibfield  {author} {\bibinfo {author} {\bibfnamefont {P.}~\bibnamefont
  {Virtanen}}, \bibinfo {author} {\bibfnamefont {R.}~\bibnamefont {Gommers}},
  \bibinfo {author} {\bibfnamefont {T.~E.}\ \bibnamefont {Oliphant}}, \bibinfo
  {author} {\bibfnamefont {M.}~\bibnamefont {Haberland}}, \bibinfo {author}
  {\bibfnamefont {T.}~\bibnamefont {Reddy}}, \emph {et~al.},\ }\bibfield
  {title} {\enquote {\bibinfo {title} {{{SciPy} 1.0: Fundamental Algorithms for
  Scientific Computing in Python}},}\ }\href
  {https://doi.org/10.1038/s41592-019-0686-2} {\bibfield  {journal} {\bibinfo
  {journal} {Nature Methods}\ }\textbf {\bibinfo {volume} {17}},\ \bibinfo
  {pages} {261--272} (\bibinfo {year} {2020})},\ \Eprint
  {https://arxiv.org/abs/1907.10121} {arXiv:1907.10121} \BibitemShut {NoStop}%
\end{thebibliography}%

\newpage
\appendix
\section{DERIVING {\boldmath $p(\chieff|q)$} FROM ALIGNED SPINS}
\label{sec:aligned-appendix}

\noindent
To derive Eq.~\eqref{eq:aligned-chieff}, first define a two-dimensional prior on $(\soz,\chieff)$ and then integrate out dependence on $\soz$.
The joint prior on $(\soz,\chieff)$ is
    \begin{equation}
    \begin{aligned}
    p(\soz,\chieff|q)
        &= p(\soz,\stz)\, \frac{\partial \stz}{\partial \chieff}\big|_{\soz} \\
        &= \left(\frac{1}{4 a_{\rm max}^2}\right) \left(\frac{1+q}{q}\right) \\
        &= \frac{1+q}{4 q a_{\rm max}^2}
    \end{aligned}
    \end{equation}
Now integrate over $\soz$ to obtain the marginal prior on $\chieff$.
To do so, though, we first need to determine the appropriate integration bounds.
In terms of $\chieff$ and $\stz$, the primary spin $\soz$ is given by
    \begin{equation}
    \soz = \chieff(1+q) - q \stz.
    \end{equation}
The maximum value $\soz$ can possibly take corresponds to the case when $\stz = -a_{\rm max}$, giving $\soz = \chieff(1+q)+ q \,a_{\rm max}$.
But we have also bounded $\soz$ itself to be less than $a_{\rm max}$.
Hence
    \begin{equation}
    \label{eq:chi-eff-upper}
    {\rm max}(\soz) = {\rm min}\Big\{\chieff(1+q)+ q \,a_{\rm max},\,\, a_{\rm max}\Big\}
    \end{equation}
Similarly,
    \begin{equation}
    {\rm min}(\soz) = {\rm max}\Big\{\chieff(1+q) - q \,a_{\rm max},\,\, -a_{\rm max}\Big\}
    \end{equation}

\noindent
\textbf{\underline{Case 1}}: Consider a case in which $\chieff(1+q)+ q \,a_{\rm max} > a_{\rm max}$.
Then ${\rm max}(\soz) = a_{\rm max}$.
Note also that
    \begin{equation}
    \begin{aligned}
    \chieff(1+q) - q \,a_{\rm max}
        &= \big(\chieff(1+q) + q \,a_{\rm max}\big) - 2q\,a_{\rm max} \\
        &> a_{\rm max} - 2q\,a_{\rm max} \\
        &> (1-2q)a_{\rm max} \\
        &\geq -a_{\rm max},
    \end{aligned}
    \end{equation}
where the last line follows from the fact that $q\leq1$.
Hence ${\rm min}(\soz) = \chieff(1+q) - q \,a_{\rm max}$, and our marginal prior on $\chieff$ is
    \begin{equation}
    \begin{aligned}
    p(\chieff|q) 
        &= \int_{\chieff(1+q) - q \,a_{\rm max}}^{a_{\rm max}} \frac{1+q}{4 q a_{\rm max}^2} ds_{1,z} \\
        &= \frac{1+q}{4 q a_{\rm max}^2} \Big( a_{\rm max} - \chieff(1+q) + q \,a_{\rm max} \Big) \\
        &= \frac{\left(1+q\right)^2 \left(a_{\rm max} - \chieff\right)}{4q a_{\rm max}^2}.
    \end{aligned}
    \end{equation}

\noindent
\textbf{\underline{Case 2}}: Next, consider the case where $\chieff(1+q) - q \,a_{\rm max} < -a_{\rm max}$, such that ${\rm min}(\soz) = -a_{\rm max}$.
This implies also that 
    \begin{equation}
    \begin{aligned}
    \chieff(1+q) + q \,a_{\rm max}
        &= \big(\chieff(1+q) - q \,a_{\rm max}\big) + 2q\,a_{\rm max} \\
        &< - a_{\rm max} + 2q\,a_{\rm max} \\
        &< (2q-1)a_{\rm max} \\
        &\leq a_{\rm max},
    \end{aligned}
    \end{equation}
so ${\rm max}(\soz) = \chieff(1+q) + q \,a_{\rm max}$.
Then the marginal prior on $\chieff$ in this case is
    \begin{equation}
    \begin{aligned}
    p(\chieff|q) 
        &= \int_{-a_{\rm max}}^{\chieff(1+q) + q \,a_{\rm max}} \frac{1+q}{4 q a_{\rm max}^2} d s_{1,z} \\
        &= \frac{1+q}{4 q a_{\rm max}^2} \Big(\chieff(1+q) + q \,a_{\rm max} + a_{\rm max} \Big) \\
        &= \frac{\left(1+q\right)^2 \left(a_{\rm max} + \chieff\right)}{4q a_{\rm max}^2}.
    \end{aligned}
    \end{equation}
    
\noindent
\textbf{\underline{Case 3}}: Finally, assume that $\chieff(1+q)+ q \,a_{\rm max} \leq a_{\rm max}$, such that ${\rm max}(\soz) = \chieff(1+q)+ q \,a_{\rm max}$.
We already covered in \textit{Case 2} the situation in which we take this maximum bound together with the minimum bound ${\rm min}(\soz) = -a_{\rm max}$, so the only unique case left to consider is one in which $\chieff(1+q)- q \,a_{\rm max} \geq - a_{\rm max}$, such that ${\rm min}(\soz) = \chieff(1+q) - q \,a_{\rm max}$.
Then
    \begin{equation}
    \begin{aligned}
    p(\chieff|q) 
        &= \int_{\chieff(1+q) - q \,a_{\rm max}}^{\chieff(1+q) + q \,a_{\rm max}} \frac{1+q}{4 q a_{\rm max}^2} ds_{1,z} \\
        &= \frac{1+q}{4 q a_{\rm max}^2} \Big(2 q \,a_{\rm max}\Big) \\
        &=\frac{1+q}{2 a_{\rm max}}.
    \end{aligned}
    \end{equation}
    
\section{DERIVING EFFECTIVE SPIN PRIORS FROM ISOTROPIC SPINS}
\label{sec:isotropic-appendix}
\subsection{COMPONENT SPIN PRIORS \boldmath ${p(s_z,s_p)}$, ${p(s_z)}$, AND ${p(s_p)}$}

\noindent
In order to obtain $p(s_z,s_p)$ from $p(a,\cos\theta) = 1/(2\amax)$, compute the Jacobian $J = \frac{\partial(a,\cos\theta)}{\partial(s_z,s_p)}$.
Written in terms of $s_z$ and $s_p$,
    \begin{equation}
    \begin{aligned}
    a &= \sqrt{s_z^2 + s_p^2} \\
    \cos\theta &= \frac{s_z}{\sqrt{s_z^2 + s_p^2}},
    \end{aligned}
    \end{equation}
and so we have
    \begin{equation}
    \begin{aligned}
    J &= \begin{vmatrix}
        \dfrac{\partial a}{\partial s_z} & \dfrac{\partial a}{\partial s_p} \\[10pt]
        \dfrac{\partial \cos\theta}{\partial s_z} & \dfrac{\partial \cos\theta}{\partial s_p}
        \end{vmatrix}  \\
    &= \begin{vmatrix}
        \dfrac{s_z}{\sqrt{s_z^2 + s_p^2}} & \dfrac{s_p}{\sqrt{s_z^2 + s_p^2}} \\
        \dfrac{s_p^2}{(s_z^2 + s_p^2)^{3/2}} & -\dfrac{s_z s_p}{(s_z^2 + s_p^2)^{3/2}}
        \end{vmatrix} \\
    &= \frac{s_p}{s_z^2 + s_p^2}
    \end{aligned}
    \end{equation}
Therefore, the joint prior on $(s_z,s_p)$ is
    \begin{equation}
    \begin{aligned}
    p(s_z,s_p)
        &= \frac{dP}{da\,d\cos\theta} \frac{\partial (a,\cos\theta)}{\partial (s_z,s_p)} \\
        &=  \frac{1}{2\amax}\frac{s_p}{s_z^2 + s_p^2}
    \end{aligned}
    \end{equation}
\\

\noindent
Next, obtain the marginal prior $p(s_z)$ by integrating over $s_p$.
Note that, since $s_z^2 + s_p^2 \leq \amax^2$, our integration bounds will run from $s_p = 0$ to $s_p = \sqrt{\amax^2 - s_z^2}$:
    \begin{equation}
    \begin{aligned}
    p(s_z)
        &= \int_0^{\sqrt{\amax^2 - s_z^2}} ds_p\, \frac{1}{2\amax}\frac{s_p}{s_z^2 + s_p^2} \\
        &= \frac{1}{2\amax} \ln\left(\frac{\amax}{|s_z|}\right).
    \end{aligned}
    \end{equation}
Similarly, we find $p(s_p)$ by integrating $s_z$ between $\pm \sqrt{\amax^2 - s_p^2}$:
    \begin{equation}
    \begin{aligned}
    p(s_p)
        &= \int_{-\sqrt{\amax^2 - s_p^2}}^{\sqrt{\amax^2 - s_p^2}} ds_z\, \frac{1}{2\amax}\frac{s_p}{s_z^2 + s_p^2} \\
        &= \frac{1}{a_{\rm max}} \cos^{-1}\left(\frac{s_p}{a_{\rm max}}\right).
    \end{aligned}
    \end{equation}
    
\subsection{EFFECTIVE ALIGNED SPIN PRIOR \boldmath ${p(\chieff|q)}$}

\noindent
We saw above that uniform and isotropic spin priors correspond to marginal priors
    \begin{equation}
    p(s_z) = \frac{1}{2 a_{\rm max}} \ln \left(\frac{a_{\rm max}}{|s_z|}\right)
    \end{equation}
on the $z$-component of each black hole's spin.
The joint prior on $\soz$ and $\stz$ is therefore
    \begin{equation}
    \begin{aligned}
    \frac{dP}{d\soz d\stz}
    &= \frac{1}{4 \amax^2} \ln \left(\frac{a_{\rm max}}{|\soz|}\right) \ln \left(\frac{a_{\rm max}}{|\stz|}\right) \\[5pt]
    &= \frac{1}{4 \amax^2} \ln \left(\frac{|\soz|}{a_{\rm max}}\right) \ln \left(\frac{|\stz|}{a_{\rm max}}\right)
    \end{aligned}
    \end{equation}
Using the definition of $\chieff$,
    \begin{equation}
    \chieff = \frac{\soz + q\stz}{1+q},
    \end{equation}
we can convert to a joint prior on $\chieff$ and $\stz$:
    \begin{equation}
    \begin{aligned}
    \frac{dP}{d\chieff d\stz}
    &= \frac{dP}{d\soz d\stz} \frac{\partial \soz}{\partial \chieff}\Big|_{\stz} \\[5pt]
    &= \frac{1+q}{4 \amax^2} \ln \left(\frac{|\soz|}{a_{\rm max}}\right) \ln \left(\frac{|\stz|}{a_{\rm max}}\right),
    \end{aligned}
    \end{equation}
where we'll now regard $\soz \equiv \soz(\chieff,\stz) = (1+q)\chieff - q\stz$ as a function of $\chieff$ and $\stz$.  \\

\noindent
We now need to integrate over $\stz$ to obtain the marginal prior on $\chieff$.
The difficult part of this is choosing appropriate integration bounds.
We have already constrained $\stz$ to run between $-\amax \leq \stz \leq \amax$.
Given  a particular $\chieff$, though, we also need to limit $\stz$ to the range where the implied $\soz(\chieff,\stz)$ is \textit{physical}.
In particular, it must be the case that
    \begin{equation}
    \begin{aligned}
    \amax &\geq \soz \\
        &\geq (1+q)\chieff -q\stz \\[10pt]
        \implies
        \stz &\geq \frac{1+q}{q}\chieff - \frac{\amax}{q}.
    \end{aligned}
    \end{equation}
Similarly, we require
    \begin{equation}
    \begin{aligned}
    -\amax &\leq \soz \\
        &\leq (1+q)\chieff -q\stz \\[10pt]
        \implies
        \stz &\leq \frac{1+q}{q}\chieff + \frac{\amax}{q}.
    \end{aligned}
    \end{equation}
So our lower and upper integration bounds are therefore
    \begin{equation}
    \begin{aligned}
    \stz^{\rm low} &= {\rm max}\Big\{ -\amax, \,\frac{1+q}{q}\chieff - \frac{\amax}{q}\Big\} \\
    \stz^{\rm high} &= {\rm min}\Big\{ \amax, \,\frac{1+q}{q}\chieff + \frac{\amax}{q}\Big\}
    \end{aligned}
    \end{equation}
\\

\noindent
Now look more closely at the two possibilities for $\stz^{\rm high}$.
We will choose $\stz^{\rm high} = \amax$ when
    \begin{equation}
    \begin{aligned}
    \amax &< \frac{1+q}{q}\chieff + \frac{\amax}{q} \\
    \implies
    \chieff &> \amax \left(\frac{q-1}{1+q}\right).
    \end{aligned}
    \label{eq:upper-bound}
    \end{equation}
Before moving any further, note that our prior on $\chieff$ must be symmetric about zero, given that our component spin priors are isotropic.
We will make our lives much easier if we leverage this symmetry and assume, for the time being, that we are working in terms of a \textit{purely positive} value of $\chieff$; i.e. the absolute value of $\chieff$.
With this in mind, we see that Eq.~\eqref{eq:upper-bound} is always satisfied: since $q<1$, the right-hand side is always negative and so always less than our purely positive $|\chieff|$.
So our upper integration bound is always
    \begin{equation}
    \stz^{\rm high} = \amax
    \end{equation}
\\

\noindent
Next let's inspect the lower integration bound.
We choose $\stz^{\rm low} = -\amax$ when
    \begin{equation}
    \begin{aligned}
    -\amax &> \frac{1+q}{q}|\chieff| - \frac{\amax}{q} \\
    \implies
    |\chieff| &< \amax \left(\frac{1-q}{1+q}\right).
    \end{aligned}
    \end{equation}
Unlike Eq.~\eqref{eq:upper-bound}, this isn't always satisfied by the positive $|\chieff|$.
So we have
    \begin{equation}
    \stz^{\rm low} = \begin{cases}
        -\amax & |\chieff| < \amax \left(\dfrac{1-q}{1+q}\right) \\[10pt]
        \dfrac{1+q}{q}|\chieff| - \dfrac{\amax}{q} & |\chieff| \geq \amax \left(\dfrac{1-q}{1+q}\right)
    \end{cases}
    \label{eq:lower-bound}
    \end{equation}
\\

\noindent
\textbf{\underline{Case 1}}: $|\chieff| < \amax \left(\dfrac{1-q}{1+q}\right)$
\\

\noindent
In this case, our marginal prior on $|\chieff|$ is
    \begin{equation}
    \begin{aligned}
    p(\chieff|q)
        &= \int_{\stz^{\rm low}}^{\stz^{\rm high}} d\stz \frac{dP}{d\chieff d\stz} \\
        &= \frac{1+q}{4 \amax^2} \int_{-\amax}^{\amax} d\stz \ln \left(\frac{|\soz|}{a_{\rm max}}\right) \ln \left(\frac{|\stz|}{a_{\rm max}}\right) \\
        &= \frac{1+q}{4 \amax^2} \int_{-\amax}^{\amax} d\stz \ln \left(\frac{\|(1+q)|\chieff| - q\stz\|}{a_{\rm max}}\right) \ln \left(\frac{|\stz|}{a_{\rm max}}\right) \\
        &= \frac{1+q}{4 \amax^2} \Bigg[
        \int_{-\amax}^0 d\stz \ln \left(\frac{\|(1+q)|\chieff| - q\stz\|}{a_{\rm max}}\right) \ln \left(\frac{-\stz}{a_{\rm max}}\right) \\&\hspace{1.5cm}
        + \int_{0}^{\amax} d\stz \ln \left(\frac{\|(1+q)|\chieff| - q\stz\|}{a_{\rm max}}\right) \ln \left(\frac{\stz}{a_{\rm max}}\right)
        \Bigg],
    \end{aligned}
    \label{eq:caseA-first}
    \end{equation}
where in the final line we've split the integration across negative and positive $\stz$ in order to resolve the absolute value $|\stz|$ appearing in the second logarithm.
In an attempt to minimize ambiguity, I use the notation $\|...|\chieff|...\|$ in cases where $|\chieff|$ is itself inside the argument of another absolute value.
\\

\noindent
In addition to $|\stz|$, we need to worry about possibly changing signs within $\|(1+q)|\chieff| - q\stz\|$ as well.
In particular, we need to know when its argument is negative and further break apart our integration bounds appropriately:
    \begin{equation}
    \begin{aligned}
    (1+q)|\chieff| - q\stz &< 0 \\[5pt]
    \implies
    \stz > \frac{1+q}{q} |\chieff|.
    \end{aligned}
    \label{eq:second-abs}
    \end{equation}
Since we chose to work with the positive quantity $|\chieff|$, we know that it is only possible for this condition to occur in the second integral of Eq.~\eqref{eq:caseA-first} where $\stz$ is positive.
If $\frac{1+q}{q}|\chieff|<\amax \implies |\chieff| < \frac{q}{1+q}\amax$, then there are places in the integrand where Eq.~\eqref{eq:second-abs} will be satisfied, and we will need to further break apart the integral to accommodate the absolute value.
If $\frac{1+q}{q}|\chieff|>\amax \implies |\chieff| > \frac{q}{1+q}\amax$, though, then we're home free.
\\

\noindent
\textbf{\underline{Case 1.A}}: $|\chieff| < \amax \left(\dfrac{1-q}{1+q}\right)$ and $|\chieff| < \amax\left(\dfrac{q}{1+q}\right)$

\noindent
In this case, we do encounter the condition in Eq.~\eqref{eq:second-abs} and we further break apart the integral:
    \begin{equation}
    \begin{aligned}
    p(\chieff|q)
    &= \frac{1+q}{4 \amax^2} \Bigg[
        \int_{-\amax}^0 d\stz \ln \left(\frac{(1+q)|\chieff| - q\stz}{a_{\rm max}}\right) \ln \left(\frac{-\stz}{a_{\rm max}}\right) \\&\hspace{1.5cm}
        + \int_{0}^{\frac{1+q}{q}|\chieff|} d\stz \ln \left(\frac{(1+q)|\chieff| - q\stz}{a_{\rm max}}\right) \ln \left(\frac{\stz}{a_{\rm max}}\right)
        \\ &\hspace{1.5cm}
        + \int_{\frac{1+q}{q}|\chieff|}^{\amax} d\stz \ln \left(\frac{q\stz - (1+q)|\chieff|}{a_{\rm max}}\right) \ln \left(\frac{\stz}{a_{\rm max}}\right)
        \Bigg].
    \end{aligned}
    \label{eq:int-temp}
    \end{equation}
\\

\noindent
This is the point where we relinquish control and turn to \textsc{Mathematica}~\cite{Mathematica}, which can eventually be coaxed into revealing
    \begin{equation}
    \begin{aligned}
    p(\chieff|q)
    &= \frac{1+q}{4 q \amax^2} \Bigg[
        q \amax \Big(
            4 + 2 \ln(\amax)
            - \ln(q^2\amax^2 - (1+q)^2 |\chieff|^2)
            \Big) \\
        &\hspace{1.3cm}
        - 2 (1+q) |\chieff|
            \tanh^{-1}\left(\frac{(1+q)|\chieff|}{q\amax}\right) \\
        &\hspace{1.3cm}
        + (1+q)|\chieff| \bigg(
            \dilog\left(\frac{-q \amax}{(1+q)|\chieff|}\right)
            - \mathrm{Re}\bigg\{ \dilog\left(\frac{q\amax}{(1+q)|\chieff|}\right)\bigg\}
            \bigg)
        \Bigg].
    \end{aligned}
    \label{eq:p-xeff-1}
    \end{equation}
Here, $\mathrm{Re}\{...\}$ denotes the real part, and $\dilog$ is the \textit{dilogarithm}, also known as Spence's function.
As discussed in Sect.~\ref{sec:isotropic-spins} above, I follow \textsc{Mathematica}'s convention in defining Spence's function, such that $\dilog(z) = \mathrm{\texttt{PolyLog[2,z]}}$.
\\

\noindent
Note that Eq.~\eqref{eq:p-xeff-1} will give divide-by-zero errors in the case that $|\chieff|=0$ exactly.
In this special case, the contribution from the second integral in Eq.~\eqref{eq:int-temp} vanishes, and we instead have
\begin{equation}
p(\chieff=0) = \frac{1+q}{2\amax} \left(2-\ln q\right)
\label{eq:p-xeff-0}
\end{equation}

\noindent
\textbf{\underline{Case 1.B}}: $|\chieff| < \amax \left(\dfrac{1-q}{1+q}\right)$ and $|\chieff| > \amax\left(\dfrac{q}{1+q}\right)$
\\

\noindent
In this case, we can happily integrate $\stz$ between $0$ and $\amax$ without worrying about the changing sign of the first logarithm:

    \begin{equation}
    \begin{aligned}
    p(\chieff|q) 
    &= \frac{1+q}{4 \amax^2} \Bigg[
        \int_{-\amax}^0 d\stz \ln \left(\frac{(1+q)|\chieff| - q\stz}{a_{\rm max}}\right) \ln \left(\frac{-\stz}{a_{\rm max}}\right) \\&\hspace{1.5cm}
        + \int_{0}^{\amax} d\stz \ln \left(\frac{(1+q)|\chieff| - q\stz}{a_{\rm max}}\right) \ln \left(\frac{\stz}{a_{\rm max}}\right)
        \Bigg] \\
    &= \frac{1+q}{4 q \amax^2} \Bigg[
        4 q \amax  + 2 q \amax \ln(\amax)
        \\& \hspace{1.5cm}
        -2(1+q) |\chieff| \tanh^{-1}\left(\frac{q\amax}{(1+q)|\chieff|}\right)
        \\[5pt]& \hspace{1.5cm}
        - q\amax \ln\left[(1+q)^2|\chieff|^2 - q^2\amax^2\right]
        \\[5pt]&\hspace{1.5cm}
        + (1+q)|\chieff| \left( \dilog\left(-\frac{q\amax}{(1+q)|\chieff|}\right)
            - \dilog\left(\frac{q\amax}{(1+q)|\chieff|}\right)
            \right)
        \Bigg].
    \end{aligned}
    \label{eq:p-xeff-2}
    \end{equation}
\\

\noindent

\noindent
\textbf{\underline{Case 2}}: $|\chieff| > \amax \left(\dfrac{1-q}{1+q}\right)$
\\

\noindent
Halfway there.
We're left to consider the second case in Eq.~\eqref{eq:lower-bound}, with $|\chieff| > \amax \left(\frac{1-q}{1+q}\right)$ and a lower integration bound $\stz^{\rm low} = \frac{1+q}{q}|\chieff| - \frac{\amax}{q}$:

    \begin{equation}
    \begin{aligned}
    p(\chieff|q)
        &= \frac{1+q}{4 \amax^2} \int_{\frac{1+q}{q}|\chieff| - \frac{\amax}{q}}^{\amax} d\stz \ln \left(\frac{|\soz|}{a_{\rm max}}\right) \ln \left(\frac{|\stz|}{a_{\rm max}}\right) \\
        &= \frac{1+q}{4 \amax^2}
        \int_{\frac{1+q}{q}|\chieff| - \frac{\amax}{q}}^{\amax} d\stz \ln \left(\frac{\|(1+q)|\chieff| - q\stz\|}{a_{\rm max}}\right) \ln \left(\frac{|\stz|}{a_{\rm max}}\right).
    \end{aligned}
    \label{eq:caseB-first}
    \end{equation}
As in \textbf{Case 1}, we can plan on splitting our integral into integration over negative and positive $\stz$, but this is only necessary when
    \begin{equation}
    \begin{aligned}
    0 &> \stz^{\rm low} \\
        &> \frac{1+q}{q}|\chieff| - \frac{\amax}{q} \\
    \implies
    |\chieff| & < \frac{\amax}{1+q}
    \end{aligned}
    \label{eq:case2-bounds}
    \end{equation}
\\

\noindent
\textbf{\underline{Case 2.A}}: $|\chieff| > \amax \left(\dfrac{1-q}{1+q}\right)$ and $|\chieff| < \dfrac{\amax}{1+q}$

\noindent
Splitting our integration across negative and positive $\stz$,
    \begin{equation}
    \begin{aligned}
    p(\chieff|q)
    &= \frac{1+q}{4 \amax^2}
        \int_{\frac{1+q}{q}|\chieff| - \frac{\amax}{q}}^{\amax} d\stz \ln \left(\frac{\|(1+q)|\chieff| - q\stz\|}{a_{\rm max}}\right) \ln \left(\frac{|\stz|}{a_{\rm max}}\right) \\
    &= \frac{1+q}{4 \amax^2} \Bigg[
        \int_{\frac{1+q}{q}|\chieff| - \frac{\amax}{q}}^{0} d\stz \ln \left(\frac{\|(1+q)|\chieff| - q\stz\|}{a_{\rm max}}\right) \ln \left(\frac{-\stz}{a_{\rm max}}\right)
        \\& \hspace{2cm}
        + \int_0^{\amax} d\stz \ln \left(\frac{\|(1+q)|\chieff| - q\stz\|}{a_{\rm max}}\right) \ln \left(\frac{\stz}{a_{\rm max}}\right)
        \Bigg]
    \end{aligned}
    \end{equation}
\noindent
We again need to worry about the absolute value in the first logarithm of our integrand, whose argument changes sign partway through integration over positive $\stz$ when $|\chieff| < \amax\left(\frac{q}{1+q}\right)$, cf. Eq.~\eqref{eq:second-abs}.
\\

\noindent
\textbf{\underline{Case 2.A.i}}: $|\chieff| > \amax \left(\dfrac{1-q}{1+q}\right)$, $|\chieff| < \dfrac{\amax}{1+q}$, and $|\chieff| < \amax \dfrac{q}{1+q}$
\\

\noindent
Splitting the integral over positive negative $\stz$ as well as positive and negative $\|(1+q)|\chieff| - q\stz\|$,
    \begin{equation}
    \begin{aligned}
    p(\chieff|q)
    &= \frac{1+q}{4 \amax^2} \Bigg[
        \int_{\frac{1+q}{q}|\chieff| - \frac{\amax}{q}}^{0} d\stz \ln \left(\frac{\|(1+q)|\chieff| - q\stz\|}{a_{\rm max}}\right) \ln \left(\frac{-\stz}{a_{\rm max}}\right)
        \\& \hspace{1.5cm}
        + \int_0^{\amax} d\stz \ln \left(\frac{\|(1+q)|\chieff| - q\stz\|}{a_{\rm max}}\right) \ln \left(\frac{\stz}{a_{\rm max}}\right)
        \Bigg] \\
    &= \frac{1+q}{4 \amax^2} \Bigg[
        \int_{\frac{1+q}{q}|\chieff| - \frac{\amax}{q}}^{0} d\stz \ln \left(\frac{(1+q)|\chieff| - q\stz}{a_{\rm max}}\right) \ln \left(\frac{-\stz}{a_{\rm max}}\right)
        \\& \hspace{1.5cm}
        + \int_0^{\frac{1+q}{q}|\chieff|} d\stz \ln \left(\frac{(1+q)|\chieff| - q\stz}{a_{\rm max}}\right) \ln \left(\frac{\stz}{a_{\rm max}}\right)
        \\& \hspace{1.5cm}
        + \int_{\frac{1+q}{q}|\chieff|}^{\amax} d\stz \ln \left(\frac{q\stz - (1+q)|\chieff|}{a_{\rm max}}\right) \ln \left(\frac{\stz}{a_{\rm max}}\right)
        \Bigg] \\
    &= \frac{1+q}{4q\amax^2} \Bigg[
        2(1+q)(\amax-|\chieff|)
            - (1+q)|\chieff| (\ln\amax)^2
        \\&\hspace{1.5cm}
        + \Big(\amax + (1+q)|\chieff| \ln[(1+q)|\chieff|]\Big)
            \ln\left(\frac{q\amax}{\amax-(1+q)|\chieff|}\right)
        \\&\hspace{1.5cm}
        -(1+q)|\chieff| \ln(\amax)
            \Big(2+\ln q-\ln[\amax-(1+q)|\chieff|]\Big)
        \\&\hspace{1.5cm}
        + q\amax \ln\left(\frac{\amax}{q \amax - (1+q)|\chieff|}\right)
        \\&\hspace{1.5cm}
        + (1+q)|\chieff| \ln\left(\frac{(\amax-(1+q)|\chieff|)(q\amax-(1+q)|\chieff|)}{q}\right)
        \\&\hspace{1.5cm}
        + (1+q)|\chieff| \left(
            \dilog\left(1-\frac{\amax}{(1+q)|\chieff|}\right)
            - \mathrm{Re}\Bigg\{
            \dilog\left(\frac{q\amax}{(1+q)|\chieff|}\right)\Bigg\}
            \right)
        \Bigg]
    \end{aligned}
    \label{eq:p-xeff-3}
    \end{equation}
\\
As a final remark on this case, the two conditions $|\chieff| < \frac{\amax}{1+q}$ and $|\chieff| < \amax \frac{q}{1+q}$ are redundant, since any $|\chieff|$ obeying the second condition will automatically obey the first.
So we can more succinctly write the conditions for this case as $|\chieff| > \amax \left(\frac{1-q}{1+q}\right)$ and $|\chieff| < \amax \frac{q}{1+q}$.

\noindent
\textbf{\underline{Case 2.A.ii}}: $|\chieff| > \amax \left(\dfrac{1-q}{1+q}\right)$, $|\chieff| < \dfrac{\amax}{1+q}$, and $|\chieff| > \amax \dfrac{q}{1+q}$
\\

\noindent
In this case, we don't need to worry about the changing sign in the first logarithm, and we only need the two terms
    \begin{equation}
    \begin{aligned}
    p(\chieff|q)
    &= \frac{1+q}{4 \amax^2} \Bigg[
        \int_{\frac{1+q}{q}|\chieff| - \frac{\amax}{q}}^{0} d\stz \ln \left(\frac{(1+q)|\chieff| - q\stz}{a_{\rm max}}\right) \ln \left(\frac{-\stz}{a_{\rm max}}\right)
        \\& \hspace{1.5cm}
        + \int_0^{\amax} d\stz \ln \left(\frac{(1+q)|\chieff| - q\stz}{a_{\rm max}}\right) \ln \left(\frac{\stz}{a_{\rm max}}\right)
        \Bigg] \\
    &= \frac{1+q}{4 q \amax^2} \Bigg[
        -|\chieff| (\ln \amax)^2 + 2(1+q)\left(
            \amax
            -|\chieff|\right)
        \\&\hspace{1.5cm}
        + q\amax \ln\left(\frac{\amax}{(1+q)|\chieff|-q\amax}\right)
        + \amax\ln\left(\frac{q\amax}{\amax-(1+q)|\chieff|}\right)
        \\&\hspace{1.5cm}
        -|\chieff|\ln(\amax) \left(
            2(1+q)
            - \ln((1+q)|\chieff|)
            -q \ln\left(\frac{(1+q)|\chieff|}{\amax}\right)
            \right)
        \\&\hspace{1.5cm}
        + (1+q)|\chieff| \ln\left(
            \frac{(-q\amax + (1+q)|\chieff|)(\amax - (1+q)|\chieff|)}
            {q}
            \right)
        \\&\hspace{1.5cm}
        +(1+q) |\chieff| \ln\left(\frac{\amax}{(1+q)|\chieff|}\right)
             \ln\left(\frac{\amax-(1+q)|\chieff|}{q}\right)
        \\&\hspace{1.5cm}
        + (1+q)|\chieff| \left(
            \dilog\left(1-\frac{\amax}{(1+q)|\chieff|}\right)
            - \dilog\left(\frac{q\amax}{(1+q)|\chieff|}\right)
            \right)
        \Bigg]
    \end{aligned}
    \label{eq:p-xeff-4}
    \end{equation}

\noindent
\textbf{\underline{Case 2.B}}: $|\chieff| > \amax \left(\dfrac{1-q}{1+q}\right)$ and $|\chieff| > \dfrac{\amax}{1+q}$
\\

\noindent
From Eq.~\eqref{eq:case2-bounds}, we know that in this case our lower integration bound $\stz^{\rm low} = \frac{1+q}{q}|\chieff| - \frac{\amax}{q}$ is greater than zero, so
    \begin{equation}
    \begin{aligned}
    p(\chieff|q)
    &= \frac{1+q}{4 \amax^2}
        \int_{\frac{1+q}{q}|\chieff| - \frac{\amax}{q}}^{\amax} d\stz \ln \left(\frac{\|(1+q)|\chieff| - q\stz\|}{a_{\rm max}}\right) \ln \left(\frac{\stz}{a_{\rm max}}\right)
    \end{aligned}
    \end{equation}
Also, as discussed for Case 2.A above, the argument of the absolute value switches sign when $|\chieff| < \frac{q\amax}{1+q}$.
Fortunately for us, this condition is never met if $|\chieff| > \frac{\amax}{1+q}$.
We therefore have the single integral
    \begin{equation}
    \begin{aligned}
    p(\chieff|q)
    &= \frac{1+q}{4 \amax^2}
        \int_{\frac{1+q}{q}|\chieff| - \frac{\amax}{q}}^{\amax} d\stz \ln \left(\frac{(1+q)|\chieff| - q\stz}{a_{\rm max}}\right) \ln \left(\frac{\stz}{a_{\rm max}}\right)
        \\
    &= \frac{1+q}{4 q\amax^2} \Bigg[
        2(1+q) (\amax-|\chieff|)
            - (1+q)|\chieff| \left(\ln \amax\right)^2
        \\ &\hspace{1.5cm}
        + \ln(\amax) \left(\amax
            -2(1+q)|\chieff|
            - (1+q)|\chieff| \ln\left(\frac{q}{(1+q)|\chieff|-\amax}\right)
            \right)
        \\ &\hspace{1.5cm}
        - \amax \ln\left(\frac{(1+q)|\chieff| - \amax}{q}\right)
        \\ &\hspace{1.5cm}
        + (1+q)|\chieff|
            \ln\left(\frac{((1+q)|\chieff|-\amax)((1+q)|\chieff|-q\amax)}{q}\right)
        \\ &\hspace{1.5cm}
        + (1+q)|\chieff|
            \ln((1+q)|\chieff|)\ln\left(\frac{q\amax}{(1+q)|\chieff|-\amax}\right)
        \\ &\hspace{1.5cm}
        - q\amax \ln\left(\frac{(1+q)|\chieff| - q\amax}{\amax}\right)
        \\ &\hspace{1.5cm}
        + (1+q)|\chieff| \left(
            \dilog\left(1-\frac{\amax}{(1+q)|\chieff|}\right)
            - \dilog\left(\frac{q\amax}{(1+q)|\chieff|}\right)
            \right)
        \Bigg]
    \end{aligned}
    \label{eq:p-xeff-5}
    \end{equation}
\\

\noindent
Finally, since $1-q$ is always less than or equal to one, the two conditions $|\chieff| > \amax \left(\frac{1-q}{1+q}\right)$ and $|\chieff| > \frac{\amax}{1+q}$ that define this case are redundant: we can replace them with the single condition $|\chieff| > \frac{\amax}{1+q}$.
    

    
\subsection{EFFECTIVE PRECESSING SPIN PRIOR \boldmath ${p(\chi_p|q)}$}

\noindent
We previously saw that, under uniform and isotropic priors on compact binary spins, the marginal prior on the in-plane spin component is
    \begin{equation}
    p(s_p) = \frac{1}{a_{\rm max}} \cos^{-1}\left(\frac{s_p}{a_{\rm max}}\right).
    \end{equation}
Using the definition of the effective precessing spin $\chi_p$,
    \begin{equation}
    \begin{aligned}
    \chi_p 
        &= \max\Big[ a_1 \sin t_1, \,\left(\frac{3+4q}{4+3q}\right) q \,a_2 \sin t_2\Big] \\
        &= \max\Big[
            \sop, \,\left(\frac{3+4q}{4+3q}\right) q \,\stp
            \Big],
    \end{aligned}
    \end{equation}
we can convolve over $\sop$ and $\stp$ to get the marginal distribution on $\chi_p$:
    \begin{equation}
    \begin{aligned}
    p(\chi_p|q)
        &= \int_0^{\amax} d\stp \int_0^{\amax} d\sop\, p(\chi_p|\sop,\stp) p(\sop) p(\stp) \\
        &= \int_0^{\amax} d\stp \int_0^{\amax} d\sop \,\delta\left(\chi_p - \max\Big[
            \sop, \,\left(\frac{3+4q}{4+3q}\right) q \,\stp
            \Big]\right) p(\sop) p(\stp).
    \end{aligned}
    \end{equation}
To handle the $\mathrm{max}[...]$ inside the delta function, we can split the integration over $\sop$ into two terms, one in which $\sop < \left(\frac{3+4q}{4+3q}\right) q \,\stp$, and the other with $\sop > \left(\frac{3+4q}{4+3q}\right) q \,\stp$.
    \begin{equation}
    \begin{aligned}
    &p(\chi_p|q) \\[10pt]
        &= \int_0^{\amax} d\stp \int_0^{\frac{3+4q}{4+3q} q \stp} d\sop \,\delta\left(\chi_p - \max\Big[
            \sop, \,\left(\frac{3+4q}{4+3q}\right) q \,\stp
            \Big]\right) p(\sop) p(\stp) \\
            &\hspace{0.5cm} +
            \int_0^{\amax} d\stp \int_{\frac{3+4q}{4+3q} q \stp}^{\amax} d\sop \,\delta\left(\chi_p - \max\Big[
            \sop, \,\left(\frac{3+4q}{4+3q}\right) q \,\stp
            \Big]\right) p(\sop) p(\stp) \\[10pt]
        &= \int_0^{\amax} d\stp \int_0^{\frac{3+4q}{4+3q} q \stp} d\sop \,\delta\left(\chi_p - \left(\frac{3+4q}{4+3q}\right) q \,\stp
            \right) p(\sop) p(\stp) \\
            &\hspace{1cm} +
            \int_0^{\amax} d\stp \int_{\frac{3+4q}{4+3q} q \stp}^{\amax} d\sop \,\delta\left(\chi_p - 
            \sop \right) p(\sop) p(\stp) \\[10pt]
        &\equiv (\mathrm{Term\,1})\, + \,(\mathrm{Term\,2}).
    \end{aligned}
    \end{equation}
Note that, since $\left(\frac{3+4q}{4+3q}\right) q \leq 1$, it will always be the case that $\left(\frac{3+4q}{4+3q}\right) q \,\stp \leq \amax$, and so we can always split the integral in this fashion.
\\

\noindent
\textbf{\underline{Term 1}}:

\noindent
Convert the delta function to a density on $\stp$.
Since
    \begin{equation}
    \begin{aligned}
    \frac{dP}{d\chi_p}
        &= \frac{dP}{d\stp}\left|\frac{d\stp}{d\chi_p}\right|
        &= \frac{dP}{d\stp}  \frac{1}{q} \frac{4+3q}{3+4q},
    \end{aligned}
    \end{equation}
the delta function can be rewritten as
    \begin{equation}
    \begin{aligned}
    \delta\left(\chi_p - \left(\frac{3+4q}{4+3q}\right)q\stp\right)
        = \delta\left(\stp - \frac{4+3q}{3+4q}\frac{\chi_p}{q}\right) \frac{1}{q} \frac{4+3q}{3+4q},
    \end{aligned}
    \end{equation}
giving
    \begin{equation}
    (\mathrm{Term\,1})
        = \frac{1}{q} \frac{4+3q}{3+4q} \int_0^{\amax} d\stp \int_0^{\frac{3+4q}{4+3q} q \stp} d\sop \,\delta\left(\stp - \frac{4+3q}{3+4q}\frac{\chi_p}{q}
            \right) p(\sop) p(\stp).
    \end{equation}
There are now two possibilities.
If $\frac{4+3q}{3+4q}\frac{\chi_p}{q} \leq \amax$, then there exists some $\stp$ that satisfies the delta function, and we have
    \begin{equation}
    \begin{aligned}
    (\mathrm{Term\,1})
        &= \frac{1}{q} \frac{4+3q}{3+4q}
            \int_0^{\chi_p} d\sop p(\sop) \,p\left(\stp =  \frac{4+3q}{3+4q}\frac{\chi_p}{q}\right) \\
        &= \frac{1}{\amax^2 q}\frac{4+3q}{3+4q}
            \left[\amax - \sqrt{\amax^2-\chi_p^2}
                + \chi_p \cos^{-1}\left(\frac{\chi_p}{\amax}\right)
                \right]
                \cos^{-1}\left(\frac{4+3q}{3+4q} \frac{\chi_p}{q\amax}\right).
    \end{aligned}
    \end{equation}
If, on the other hand, $\frac{4+3q}{3+4q}\frac{\chi_p}{q} > \amax$, then there is no $\stp$ satisfying the delta function, and our integral evaluates to zero.
So
    \begin{equation}
    (\mathrm{Term\,1}) = \begin{cases}
        \begin{aligned}
        &\frac{1}{\amax^2 q}\frac{4+3q}{3+4q}
        \cos^{-1}\left(\frac{4+3q}{3+4q} \frac{\chi_p}{q \amax}\right) \\
        &\hspace{0.5cm}\times
            \left[\amax - \sqrt{\amax^2-\chi_p^2}
                + \chi_p \cos^{-1}\left(\frac{\chi_p}{\amax}\right)
                \right]
        \end{aligned}
            & \left(\chi_p \leq \frac{3+4q}{4+3q} q \amax\right) \\[40pt]
        0 & \left(\chi_p > \frac{3+4q}{4+3q} q \amax\right)
    \end{cases}
    \label{eq:chip-1}
    \end{equation}
\\

\noindent
\textbf{\underline{Term 2}}:
In this case, trivially convert the delta function  into a density on $\sop$:
    \begin{equation}
    \begin{aligned}
    (\mathrm{Term}\,2)
        &= \int_0^{\amax} d\stp \int_{\frac{3+4q}{4+3q} q \stp}^{\amax}
            d\sop \,\delta\left(\chi_p - 
            \sop \right) p(\sop) p(\stp) \\
        &= \int_0^{\amax} d\stp \int_{\frac{3+4q}{4+3q} q \stp}^{\amax}
            d\sop \,\delta\left(\sop - \chi_p \right) p(\sop) p(\stp).
    \end{aligned}
    \end{equation}
In order for the delta function to be non-zero, we need our lower integration bound on $\sop$ to satisfy $\sop^{\rm low} = \frac{3+4q}{4+3q} q \stp \leq \chi_p$.
Rearranging, this means that we should impose the bound of $\stp \leq \frac{4+3q}{3+4q} \frac{\chi_p}{q}$.
The upper integration bound on $\stp$ is therefore
    \begin{equation}
    \stp^{\rm high} = {\rm min}
        \left[\frac{4+3q}{3+4q} \frac{\chi_p}{q},\amax\right].
    \end{equation}
\\

\noindent
First consider the case that $\chi_p \leq \frac{3+4q}{4+3q} q \amax$, such that $\stp^{\rm high} = \frac{4+3q}{3+4q} \frac{\chi_p}{q}$:
    \begin{equation}
    \begin{aligned}
    (\mathrm{Term}\,2)
        &= \int_0^{\frac{4+3q}{3+4q} \frac{\chi_p}{q}} d\stp \int_{\frac{3+4q}{4+3q} q \stp}^{\amax}
            d\sop \,\delta\left(\sop - \chi_p \right) p(\sop) p(\stp) \\[10pt]
        &= \frac{1}{\amax} \cos^{-1}\left(\frac{\chi_p}{\amax}\right)
        \int_0^{\frac{4+3q}{3+4q} \frac{\chi_p}{q}} d\stp \,\frac{1}{\amax} \cos^{-1}\left(\frac{\stp}{\amax}\right) \\[10pt]
        &= \frac{1}{\amax^2 q}\frac{4+3q}{3+4q}
            \cos^{-1}\left(\frac{\chi_p}{\amax}\right)
            \bigg[
                \amax q \frac{3+4q}{4+3q} 
                \\ &\hspace{2cm}
                - \sqrt{\amax^2 q^2 \left(\frac{3+4q}{4+3q}\right)^2
                    -\chi_p^2}
                + \chi_p \cos^{-1}\left(\frac{4+3q}{3+4q}
                    \frac{\chi_p}{\amax q}\right)
            \bigg]
    \end{aligned}
    \end{equation}
\\

\noindent
Meanwhile, if $\chi_p > \frac{3+4q}{4+3q} q \amax$, we integrate up to $\stp = \amax$, giving
    \begin{equation}
    \begin{aligned}
    (\mathrm{Term}\,2)
        &= \int_0^{\amax} d\stp \int_{\frac{3+4q}{4+3q} q \stp}^{\amax}
            d\sop \,\delta\left(\sop - \chi_p \right) p(\sop) p(\stp) \\[10pt]
        &= \frac{1}{\amax} \cos^{-1}\left(\frac{\chi_p}{\amax}\right)
        \int_0^{\amax} d\stp \,\frac{1}{\amax} \cos^{-1}\left(\frac{\stp}{\amax}\right) \\[10pt]
        &= \frac{1}{\amax} \cos^{-1}\left(\frac{\chi_p}{\amax}\right).
    \end{aligned}
    \end{equation}
Together,
    \begin{equation}
    (\mathrm{Term\,2}) = \begin{cases}
        \begin{aligned}
        &\frac{1}{\amax^2 q}\frac{4+3q}{3+4q}
            \cos^{-1}\left(\frac{\chi_p}{\amax}\right)
            \bigg[
                \amax q \frac{3+4q}{4+3q} 
                \\[5pt] &\hspace{2cm}
                - \sqrt{\amax^2 q^2 \left(\frac{3+4q}{4+3q}\right)^2
                    -\chi_p^2}
                \\[5pt] &\hspace{2cm}
                + \chi_p \cos^{-1}\left(\frac{4+3q}{3+4q}
                    \frac{\chi_p}{\amax q}\right)
            \bigg]
        \end{aligned}
            \qquad & \left(\chi_p \leq \frac{3+4q}{4+3q} q \amax\right) \\[55pt]
        \dfrac{1}{\amax} \cos^{-1}\left(\dfrac{\chi_p}{\amax}\right)
        & \left(\chi_p > \frac{3+4q}{4+3q} q \amax,\,\, \chi_p<\amax\right) \\[20pt]
        0 & \Big( \chi_p \geq \amax\Big)
    \end{cases}
    \label{eq:chip-2}
    \end{equation}
\\

\noindent
With both Term 1 and Term 2 in hand, the full marginal prior $p(\chi_p|q)$ is now just the sum of Eqs.~\eqref{eq:chip-1} and \eqref{eq:chip-2}!

\end{document}